\documentclass[prd,aps,notitlepage,superscriptaddress,nofootinbib]{revtex4-1}
\usepackage{amsmath}
\usepackage{slashed}
\usepackage{epsfig}
\usepackage{color}
\def\bea#1\eea{\begin{align}#1\end{align}} 
\newcommand{\nnu}{\nonumber\\}
\newcommand{\bef}{\begin{figure}[htb]\centering}
\newcommand{\eef}{\end{figure}}
\allowdisplaybreaks

\begin{document}
\title{Semi-Inclusive Jet Functions and Jet Substructure in $J_{E_T}^{(I)}$ and $J_{E_T}^{(II)}$ Algorithms}
\author{Lei Wang}
\affiliation{Key Laboratory of Quark and Lepton Physics (MOE) and Institute of Particle Physics, Central China Normal University, Wuhan 430079, China}

\author{Zhong-Bo Kang}
\affiliation{Department of Physics and Astronomy,University of California, Los Angeles, California 90095,USA}
\affiliation{Mani L. Bhaumik Institute for Theoretical Physics, University of California, Los Angeles, California 90095,USA}

\author{Hongxi Xing}
\affiliation{Guangdong Provincial Key Laboratory of Nuclear Science, Institute of Quantum Matter, South China Normal  University, Guangzhou, 510006, China  }

\author{Ben-Wei Zhang}
\affiliation{Key Laboratory of Quark and Lepton Physics (MOE) and Institute of Particle Physics, Central China Normal University, Wuhan 430079, China}
\affiliation{Guangdong Provincial Key Laboratory of Nuclear Science, Institute of Quantum Matter, South China Normal  University, Guangzhou, 510006, China  }

\begin{abstract}
Within the framework of Soft Collinear Effective Theory, we present calculations of semi-inclusive jet functions and fragmenting jet functions at next-to-leading order (NLO) for both quark- and gluon-initiated jets, for jet algorithms of $J_{E_T}^{(I)}$ and $J_{E_T}^{(II)}$ where one maximizes a suitable jet function. We demonstrate the consistency of the obtained results with the standard perturbative QCD calculations for $J_{E_T}^{(I)}$ algorithm, while the results for fragmenting jet functions with the $J_{E_T}^{(II)}$ algorithm are new. The renormalization group (RG) equation for both semi-inclusive jet functions and fragmenting jet functions are derived and shown to follow the time-like DGLAP evolution equations, independent of specific jet algorithms. The RG equation can be used to resum single logarithms of the jet size parameter $\beta$ for highly collimated jets in these algorithms where $\beta \gg 1$. 
\end{abstract}
\maketitle

\section{Introduction}
In high energy proton-proton and nucleus-nucleus collisions, tremendous amount of collimated jets of hadrons are produced and measured at the Large Hadron Collider (LHC). The studies of the production rate of these jets and their substructures emerged as essential tools to probe the fundamental properties of Quantum Chromodynamics (QCD)~\cite{Sterman:1977wj,Ellis:2007ib,Sapeta:2015gee,Buttar:2008jx,Salam:2009jx,Altheimer:2012mn} and nucleon structure~\cite{Lai:1996mg,Martin:2001es,Aschenauer:2019uex,Arratia:2019vju}. In addition, they are also involved in searching for signals of new physics beyond the standard model~\cite{Stump:2003yu,Butterworth:2008iy}, as well as in identifying the properties of the hot dense medium, quark-gluon plasma, created in heavy ion collisions~\cite{Vitev:2008rz,Vitev:2009rd,Muller:2012zq,Armesto:2015ioy,Connors:2017ptx}. For some recent review, see Refs.~\cite{Larkoski:2017jix,Wang:2016opj,Page:2019gbf} and references therein.

Because of the crucial roles of jets, significant theoretical and experimental efforts have been devoted into the study of jets in both particle and nuclear physics communities. For example, within the framework of Soft Collinear Effective Theory (SCET)~\cite{Bauer:2000ew,Bauer:2000yr,Bauer:2001ct,Bauer:2001yt}, the cross section for inclusive jet production in $pp$ collisions can be factorized into a convolution product of initial-state parton distributions functions (PDFs) $f$, hard-part coefficient $H$ and final state semi-inclusive jet functions (siJFs)~$J$~\cite{Kang:2016mcy,Dai:2016hzf,Kaufmann:2015hma}
\bea
d\sigma^{pp\to jet +X} \sim \sum_{i,j,k} f_i\otimes f_j \otimes H_{ij}^k\otimes J_k.
\label{eq-fac}
\eea
The semi-inclusive jet functions $J_k$ characterize the probability density of a parton $k$ that is transformed to a jet. Similarly, one can also study the internal structure of the jet by measuring, e.g. the distribution of identified hadrons inside the jet, which is described by the so-called semi-inclusive fragmenting jet functions (siFJFs) $\mathcal{G}_k^{h}$~\cite{Kang:2016ehg}. The relevant factorization formula is very similar to that in Eq.~\eqref{eq-fac}, but replacing $J_k \to \mathcal{G}_k^{h}$~\cite{Kang:2016ehg,Dai:2016hzf}. Based on this factorization formula, significant extensions and improvements are established including the study of jet quenching physics for light flavors~\cite{Kang:2017frl}, next-to-leading order (NLO) calculations for heavy flavor jet in vacuum~\cite{Fickinger:2016rfd,Dai:2018ywt} and in  medium~\cite{Li:2018xuv}, as well as the application to study heavy quarkonium production mechanism~\cite{Kang:2017yde,Bain:2017wvk}. 

In all these works, typical algorithms, for example, cone and/or anti-$k_T$ algorithm, are used to match to the experimental analysis. The final results at NLO show single logarithmic structure $\alpha_s^{n}\ln^{n}R$ with $R$ represents the size of the identified jets. These logarithms spoil the convergence of perturbative expansion, and thus need to be resummed to all orders. This can be realized through the renormalization group (RG) equations of the relevant semi-inclusive jet functions. It has been shown that such RG equations are the same as the time-like DGLAP evolution equations~\cite{Kang:2016mcy,Dai:2016hzf,Dasgupta:2014yra}.

A while ago, another type of jet algorithm is proposed initially by Georgi~\cite{Georgi:2014zwa}, where the idea is to cluster jets by maximizing a fixed/suitable function of the total four-momentum of jets. This algorithm was subsequently improved by Bai et al.~\cite{Bai:2014qca} by using the total transverse energy instead of the energy in the fixed function, which is more appropriate for hadronic collisions such as those at the LHC, because transverse energies are boost invariant. This type of algorithm has been implemented into the standard pQCD calculations at NLO for single-inclusive jet production~\cite{Kaufmann:2014nda} and jet fragmentation functions~\cite{Kaufmann:2015hma} at hadron colliders, in which infrared safety of the algorithms is established and comparisons to cone and anti-$k_T$ algorithms are presented. Since these algorithms maximize a suitable jet functions, we will refer to them as ``{\it maximized jet algorithms}'' for simplicity. 

In this work, within the framework of SCET, we perform explicit calculations at NLO to study siJFs and siFJFs in maximized jet algorithms of $J_{E_T}^{(I)}$ and $J_{E_T}^{(II)}$~\cite{Bai:2014qca} as defined in the next section. We demonstrate the consistency of our results with the standard perturbative QCD calculations for $J_{E_T}^{(I)}$ algorithm~\cite{Kaufmann:2014nda,Kaufmann:2015hma}. We also perform the calculations in both light-cone gauge and covariant gauge for cross-checking our results, while the previous calculations are typically performed in covariant gauge, see e.g.~\cite{Ellis:2010rwa,Kang:2016mcy}. For both siJFs and siFJFs, we find exactly the same divergent behavior, which leads to exactly the same RG equations, i.e. time-like DGLAP evolution equations, independent of specific jet algorithms, while the remaining finite parts exhibit the algorithm dependence. These RG equations can be used to perform the resummation of single logarithms of the jet size parameter $\beta$ for maximized jet algorithms, where $\beta\gg 1$ corresponds to highly collimated jets. In this sense, the situation is very similar to the case in the anti-$k_T$ algorithm, where the RG equations are used to resum single logarithms of the jet radius $R$ for the narrow jets with $R\ll 1$~\cite{Kang:2016mcy}. Note that we focus on fully analytical calculations of the siJFs and siFJFs at NLO in the current work, and we leave phenomenological implementations of these results in pp and AA collisions for future publications.  

The remainder of this paper is organized as follows. In section~\ref{sec-definition}, we recall the operator definition of siJFs and give introduction to the maximized jet algorithms. In section~\ref{sec-jet}, we present explicit calculations of siJFs for quark and gluon jets at NLO by considering both $J_{E_T}^{(I)}$ and $J_{E_T}^{(II)}$ maximized jet algorithms and compare them with the standard pQCD results in the literature. In section~\ref{sec-fjf}, we extend the calculation to semi-inclusive fragmenting jet functions at NLO. We conclude our paper in section~\ref{sec-sum}.

\section{Definitions and Maximized jet algorithms}
\label{sec-definition}
In this section we start by giving the definition of the semi-inclusive quark and gluon jet functions in SCET, which can be constructed from the corresponding gauge invariant quark and gluon fields. The siJFs are interpreted as the probability density of the parton to transform into a jet. In light-cone coordinates, they are given by the following operator definitions~\cite{Kang:2016mcy} for the quark and gluon jets, respectively 
\bea
J_q(z, E_{\rm J}) =& \frac{z}{2N_c} {\rm Tr} \left[\frac{{\slashed{\bar n}}}{2} \langle 0|\delta(\omega - \bar{n} \cdot \mathcal{P}){\chi}_n(0)|JX\rangle \langle JX|\bar{{\chi}}_n(0)|0\rangle \right], 
\\
J_g(z, E_{\rm J}) =& -\frac{z\omega}{2(N_c^2 -1 )} \langle 0|\delta(\omega - \bar{n} \cdot \mathcal{P}) \mathcal{B}_{n\perp \mu}(0)|JX\rangle \langle JX|\mathcal{B}^\mu _{n \perp}(0)|0\rangle,
\eea
where $E_{\rm J} = z E$ is the jet energy, and $E$ is the energy of the parton initiating the jet, $\mathcal{P}$ is the label momentum operator, and the state $|JX\rangle$ represents the final-state observed jet $J$ and unobserved particles $X$. $\chi_n$ and $\mathcal{B}_{n\perp \mu}$ are gauge invariant $n$-collinear quark and gluon fields, respectively. Note that the light-cone vector $n^{\mu}$ is defined along the jet axis, and its conjugate vector is $\bar n^{\mu}$. In the frame where the jet has no transverse momentum, we can write $n^\mu = (1, 0, 0, 1)$ and $\bar n^{\mu}=(1,0,0,-1)$, which satisfies $n^2=\bar n^2 =0$ and $n\cdot \bar n = 2$. In such a frame, we define $\omega = \bar n\cdot P$ and $\omega_{\rm J} = \bar n\cdot P_J$ as the large light-cone components of the momenta for the parton initiating the jet ($P$) and the jet itself ($P_J$), respectively. For a collimated jet, we have $\omega\approx 2 E$ and $\omega_{\rm J} \approx 2 E_{\rm J}$. 

In this work, we neglect the nonperturbative hadronization effect and consider only the perturbative aspect of the jet functions, thus one can calculate both quark and gluon initiated jets perturbatively. At leading order (LO) in which one parton forms the jet, the jet functions are independent of specific jet algorithms and are simply delta functions
\bea
J_q^{(0)}(z, E_{\rm J}) = J_g^{(0)}(z, E_{\rm J}) = \delta(1-z),
\eea
where the superscript $(0)$ denotes the LO result.

At next-to-leading order (NLO), one has to consider the phase space constraints for the radiated parton according to specific jet clustering algorithms. It is this constraint that leads to the algorithm dependence of the jet functions. There are so far two classes of broadly defined algorithms: cone algorithms~\cite{Blazey:2000qt} and successive recombination algorithms~\cite{Ellis:1993tq}. Cone algorithms include the Snowmass and SIS cone algorithms~\cite{Salam:2007xv}, while recombination algorithms include the Cambridge-Aachen, $k_T$  and anti-$k_T$ algorithms~\cite{Dokshitzer:1997in,Cacciari:2008gp}. In Refs.~\cite{Kang:2016mcy,Kang:2017mda}, the semi-inclusive jet functions in cone and anti-$k_T$ algorithms have been calculated up to NLO. 
In the present paper, we will extend the calculation to two other jet finding methods $J_{E_T}^{(I)}$  and  $J^{(II)}_{E_T}$~\cite{Georgi:2014zwa,Bai:2014qca}, in which the primary idea is based on maximizing a fixed function of the total four-momentum of the final-state particles. Following Ref.~\cite{Georgi:2014zwa}, one defines one such function -- the $J_{E_T}^{(I)}$ function -- as follows 
\begin{eqnarray}
J_{E_T}^{(I)}\left(P_{\rm set}^{\mu}\right) \equiv E_{\rm set}-\beta \frac{m_{\rm set}^{2}}{E_{\rm set}} = E_{\rm set}\left[ 1 - \beta \frac{m_{\rm set}^{2}}{E_{\rm set}^2}\right].
\label{eq:jet_define_E}
\end{eqnarray}
Here, $E_{\rm set}$ and $P_{\rm set}^{\mu}$ is the total energy and four-momentum of a given subset of the final-state particles, and $m_{\rm set}$ is its invariant mass, i.e. $P_{\rm set}^2 = m_{\rm set}^2$. Ref.~\cite{Bai:2014qca} improves the above definition by using the transverse energy defined as $(E_{\rm set}^{\perp})^{2} \equiv P_{\perp}^2+m_{\rm set}^{2}$ instead of the energy $E_{\rm set}$, where $P_\perp$ is the magnitude of the transverse momentum. Thus we have the $J_{E_T}^{(I)}$ function as
\begin{eqnarray}
J_{E_{T}}^{(I)}\left(P_{\rm set}^{\mu}\right) \equiv E_{\rm set}^{\perp} \left[1 -\beta \frac{m_{\rm set}^{2}}{\left(E_{\rm set}^{\perp}\right)^2}\right].
\label{eq-jet_define}
\end{eqnarray}
Apparently the definition in Eq.~\eqref{eq:jet_define_E} is more suitable for jet production in $e^+e^-$ collisions where the energy of the jet is relevant. On the other hand, the definition in Eq.~\eqref{eq-jet_define} uses transverse energies which are boost-invariant and hence more suitable for the application to hadronic scattering, such as $pp$ collisions at the LHC. Note that for collimated jets, so-called narrow jet approximation~(NJA) applies, and one could replace the transverse energies $E_\perp$ by the transverse momenta $P_\perp$. In our calculations below, we choose a frame in which the jet has zero transverse momentum, and thus we will follow Eq.~\eqref{eq:jet_define_E} in most of our calculations. However, once our calculations is done, translating from Eqs.~\eqref{eq:jet_define_E} to \eqref{eq-jet_define} will correspond to simply replacing the jet energy $E_{\rm J}$ in our final expressions by the jet transverse momentum $P_{\rm J\perp}$ for studying jet production in $pp$ collisions at the LHC. 

For this new jet algorithm, the parameter $\beta \ge 1$ specifies the algorithm and is introduced to determine the geometric size of the jet. By maximizing the $J_{E_T}^{(I)}$ function in Eq.~\eqref{eq-jet_define}, the final state particles are forced into the collimated jets. For example, if the invariant mass $m_{\mathrm{J}}$ is large, the set will fail to produce a global maximum of $J_{E_T}^{(I)}$. This means only a subset that has large transverse energy but small invariant mass can form the jet. A reconstructed jet thus maximizes the function  $J_{E_T}^{(I)}$ with the value
\bea
J_{E_{T}}^{(I)}\left(P_{\rm J}^{\mu}\right) \equiv E_{\rm J}^{\perp} \left[1 -\beta \frac{m_{\rm J}^{2}}{\left(E_{\rm J}^{\perp}\right)^2}\right]\,.
\eea
In other words, when the jet is formed, the corresponding total four-momentum of the subset of the final-state particles $P_{\rm set}$ gives the jet momentum $P_{\rm J}$, and $E_{\rm J}^\perp$ and $m_{\rm J}$ are the transverse energy and invariant mass of the jet. The algorithm is iterative, i.e., once a jet has been found, the algorithm proceeds by removing the subset from the list of particles in the event, and apply iteratively to the remaining ones.

One may further vary the function $J_{E_T}^{(I)}$ by changing the weighted functions~\cite{Kaufmann:2014nda}. For instance, one can define the $J_{E_{T}}^{(n)}$ algorithm as follows
\begin{eqnarray}
J_{E_{T}}^{(n)}\left(P_{\mathrm{J}}^{\mu}\right) \equiv  (E_{\rm set}^\perp)^{n} \left[1-\beta \frac{m_{\rm set}^{2}}{(E_{\rm set}^{\perp})^{2}}\right],
\end{eqnarray}
where $n=1$ and $n=2$ correspond to the $J_{E_{T}}^{(I)}$ and $J_{E_{T}}^{(II)}$ algorithms~\footnote{Technically speaking, the above $J_{E_{T}}^{(II)}$ only corresponds to a special (one-prong) case of the original definition given in~\cite{Bai:2015fka}. This is sufficient for our purpose, where we focus on inclusive jet production and hadron distribution inside the jet and  one-prong/two-prong distinction of the jet substructure is not needed.}, respectively. In the present paper, we will focus on $J_{E_T}^{(I)}$ and  $J_{E_{T}}^{(II)}$ algorithms to explicitly show the algorithm dependence of siJFs. We will derive the NLO siJFs by considering highly collimated jets, i.e., $\beta \gg 1$. In this case, one can take the so-called narrow jet approximation and obtain fully analytical expressions.

\section{The Semi-inclusive  jet functions in maximized algorithm}
\label{sec-jet}
In this section, we present the detailed calculations at NLO for semi-inclusive jet functions for both quark and gluon jets in the maximized jet algorithm. As we have already mentioned, we perform the calculations in both light-cone gauge and covariant gauge for cross-checking our results, while the previous calculations are typically performed in covariant gauge, see e.g.~\cite{Ellis:2010rwa,Kang:2016mcy}. We denote the incoming parton with momentum $\ell=(\ell^-=\omega, \ell^+, 0_{\perp} )$ splits into two partons with one of them carry momentum $q= (q^-,q^+,q_\perp)$.

\subsection{The semi-inclusive quark jet function }

For quark jet function, the total contributions from the relevant diagrams, as shown in Fig.~\ref{fig-quark}, have been written down explicitly in $d=4-2\epsilon$ dimensions in Ref.~\cite{Kang:2016mcy}. We rewrite it here for the completeness of showing the calculation
\bea
J_q(z, E_{\rm J}) =& g_s^2 \left( \frac{\mu^2 e^{\gamma_E}}{4\pi}\right)^\epsilon C_F \int \frac{d\ell^+}{2\pi} \frac{1}{\ell^+} \int\frac{d^d q}{(2\pi)^d} \left[4\frac{\ell^+}{q^-} + 2(1-\epsilon) \frac{\ell^+ - q^+}{\omega-q^-}\right]2\pi \delta (q^+q^- - q_\perp^2 )
\nnu
&\times  2\pi \delta \left(\ell^+ - q^+ - \frac{q_\perp^2}{\omega-q^-}\right)\delta\left(z-\frac{\omega_{\rm J}}{\omega}\right)
\Theta(q^-) \Theta(q^+) \Theta(\omega-q^-) \Theta(\ell^+ - q^+) \Theta_{\rm alg},
\label{eq-Jq}
\eea
In this case, $q$ denotes the momentum for the radiated gluon, $\Theta_{\rm alg}$ is determined by the jet algorithm and by the kinematics of the radiated parton whether it is inside the jet. As shown in Fig.~\ref{fig-quark}, there are three situations that we need to take into consideration. The diagram (A) is for the case that both quark and gluon are inside the jet, diagrams (B) and (C) are for only quark is inside the jet, and only gluon is inside the jet, respectively. Each case has different $\Theta_{\rm alg}$ to be specified below, the combination of these three cases gives the final result.
\bef
\includegraphics[width = 0.6\linewidth]{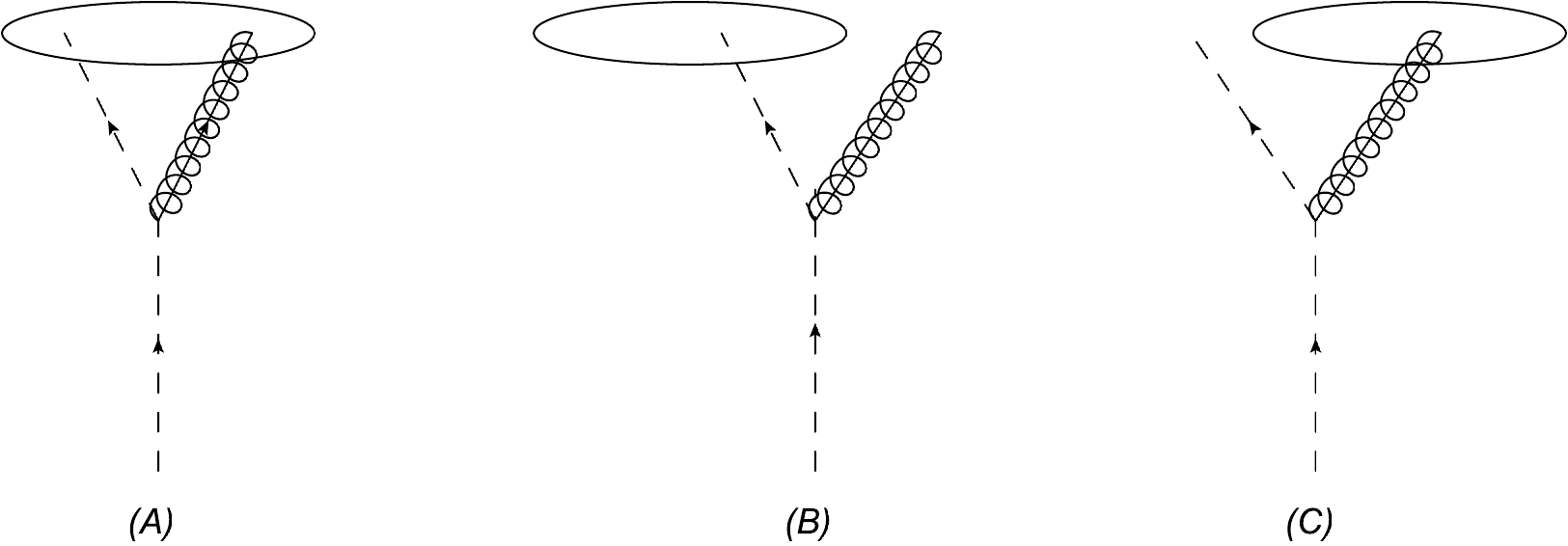}
\caption{The three situations that contribute to the semi-inclusive quark jet function: (A) both quark and gluon are inside the jet, (B) only quark is inside the jet, and (C) only gluon is inside the jet.}
\label{fig-quark}	
\eef

In the situation that both quark and gluon are inside the jet, the incoming quark energy $E$ is all converted to the jet energy $E_{\mathrm{J}}$, which leads to $z=E_{\mathrm{J}}/E=1$. 
In this case, we can perform the integration over quantities $\ell^+$ and $q^+$ in Eq.~(\ref{eq-Jq}), thus the quark jet function becomes,
\bea
J_{q\to qg}(z, E_{\rm J})=\delta(1-z) \frac{\alpha_s}{\pi}  \frac{(\mu^2 e^{\gamma_E})^\epsilon}{\Gamma(1-\epsilon)} \int_{0}^{1} dx \hat{P} _{qq}(x,\epsilon)[x(1-x)]^{-\epsilon}\int \frac{dm^2}{(m^2)^{1+\epsilon}} \Theta_{\rm alg},
\label{eq:siJFs}
\eea
where the subscript ``$q\to qg$" stands for the situation that both the radiated quark and gluon are inside the jet, $x = (\ell^- - q^-)/\ell^-$ is the momentum fraction of the initial quark carried by the final-state quark, and $m^2=\ell^2$ is the invariant mass for the final-state parton pair, i.e. $q+g$. Of course, since $q$ and $g$ together forms the jet, $m^2 = m_{\rm J}^2$ is just the jet invariant mass, and it is related to the radiated gluon transverse momentum $q_{\perp}$ by $m^2 = \frac{{q}_\perp^2}{x(1-x)}$. The functions $\hat{P}_{ij} (x,\epsilon)$ are
\bea
\hat{P}_{q q}(x, \epsilon)&=C_{F}\left[\frac{1+x^{2}}{1-x}-\epsilon(1-x)\right],
\label{Pqqhat}
\\
\hat{P}_{g q}(x, \epsilon)&=C_{F}\left[\frac{1+(1-x)^{2}}{x}-\epsilon x\right],
\label{Pgqhat}
\\
\hat{P}_{q g}(x, \epsilon)&=T_{F}\left[1-\frac{2 x(1-x)}{1-\epsilon}\right],
\label{Pqghat}
\\
\hat{P}_{g g}(x, \epsilon)&=C_{A}\left[\frac{2x}{1-x} + \frac{2(1-x)}{x} +2x(1-x)\right].
\label{Pgghat}
\eea

In the channel $q\to qg$, to make sure that the final $q$ and $g$ actually forms one jet, one has to require that the value of the $J_{E_T}^{(n)}$ function constructed from the two partons together is larger than the $J_{E_T}^{(n)}$ value constructed for each parton individually. When both $q$ and $g$ forms the jet, we have 
\bea
J_{E_T}^{(n)}(q+g) = E^n \left[1-\beta \frac{m^{2}}{E^2}\right],
\label{eq:q+g}
\eea
where we remind that $E$ is the energy of the incoming quark and is the same as the jet energy $E_{\rm J} = E$ when $q$ and $g$ together forms the jet. On the other hand, for the case in which either $q$ or $g$ forms the jet, we have
\bea
J_{E_T}^{(n)}(q) = (E_{q})^n,
\qquad
J_{E_T}^{(n)}(g) = (E_{g})^n,
\label{eq:qg}
\eea
where we have used the fact that the invariant mass of the final-state quark $q$ or $g$ vanishes. Thus the requirements of the maximized jet algorithm $J_{E_T}^{(n)}(q+g) \geq J_{E_T}^{(n)}(q)$ and $J_{E_T}^{(n)}(q+g) \geq J_{E_T}^{(n)}(g)$ lead to the following constraint, 
\begin{eqnarray}
E^n \left[1-\beta \frac{m^{2}}{E^2}\right] \geq \max \left[(E_{q})^n, \, (E_{g})^n \right]\,.
\end{eqnarray}
Realizing $E_q = x E$ and $E_g = (1-x) E$, we obtain the following constraint for $m^2$,
\bea
m^2 \leq \frac{E^{2}}{\beta} \min \left[1-(1-x)^{n},\, 1-x^{n}\right]\,.
\label{eq-constraint1}
\eea
This leads to the following algorithm constraint
\bea
\Theta_{\rm alg} = \Theta\left[\frac{E^{2}}{\beta} \min \left(1-(1-x)^{n}, 1 - x^{n}\right) 
- m^{2} \right],
\label{eq-constraint2}
\eea
where $\Theta$ is the step function. 

Let us first consider the $J_{E_T}^{(I)}$ algorithm, the calculation of the $J_{E_T}^{(II)}$ algorithm will follow the same procedure. Apply the above constraint to the jet function with $n=1$, we arrive at 
\begin{eqnarray}
J_{q\rightarrow qg}^{(I)}(z, E_{\rm J}) =\delta(1-z) \frac{\alpha_s}{2\pi} \frac{(\mu^2 e^{\gamma_E})^\epsilon}{\Gamma(1-\epsilon)} \left(-\frac{1}{\epsilon}\right)  \left(\frac{E_{\rm J}^2}{\beta} \right)^{-\epsilon} I_{qq}^{(1)},
\label{eq-J1}
\end{eqnarray}
where 
\begin{eqnarray}
I_{qq}^{(1)} \equiv\left[\int_{0}^{1 / 2} d x\left(1-(1-x)\right)^{-\epsilon}+\int_{1 / 2}^{1} d x\left(1-x\right)^{-\epsilon}\right] x^{-\epsilon}(1-x)^{-\epsilon} \hat P_{qq}(x).
\end{eqnarray}
 Expand $I_{qq}^{(1)}$ in $\epsilon$, we get the explicit expression
\begin{eqnarray}
I_{q q}^{(1)}=C_{F}\left[-\frac{1}{\epsilon}-\frac{3}{2}+\epsilon\left(-5+\frac{\pi^{2}}{2}-\frac{3}{2} \ln 2\right)\right].
\end{eqnarray}
Substituting the above expression to Eq.~(\ref{eq-J1}), we obtain the quark jet function when both quark and gluon are inside the jet for the $J_{E_T}^{(I)}$ algorithm
\begin{eqnarray}
J_{q\rightarrow qg}^{(I)}(z, E_{\mathrm{J}}) = \delta(1-z) \frac{\alpha_s}{2\pi} C_F\left(\frac{1}{\epsilon^2} + \frac{3}{2\epsilon} + \frac{1}{\epsilon} \hat{L}+\frac{1}{2}\hat{L}^2 +\frac{3}{2} \hat{L} +5+\frac{3}{2} \ln 2 - \frac{7}{12}\pi^2  \right),
\label{eq-q2qg}
\end{eqnarray}
with $\hat{L}$ defined as 
\begin{eqnarray}
\hat{L}=\ln \left(\frac{\beta \mu^2}{E_{\mathrm{J}}^{2}}\right).
\end{eqnarray}
For highly collimated jet, i.e., $\beta\gg 1$, the above logarithmic term becomes very large and needs to be resummed. This is similar to that in $k_t$-type algorithms~\cite{Kang:2016mcy}, where large logarithmic terms of $R$ needs to be resummed for small jet radius $R$. 

Now let's consider the situation that only the final state quark forms the jet. The corresponding diagram is presented in Fig.~\ref{fig-quark}(B). In this case, the final-state quark forms the jet with jet energy $E_{\mathrm{J}}=z E$, namely only a fraction $z$ of the initial parton energy $E$ falls inside the jet. 
Follow the same calculation as before, we obtain
\begin{eqnarray}
J_{q\rightarrow q(g)} (z, E_{\mathrm{J}}) =  \frac{\alpha_s}{2\pi} \frac{(\mu^2 e^{\gamma_E})^\epsilon}{\Gamma(1-\epsilon)} \hat{P}_{qq}(z,\epsilon) [z(1-z)]^{-\epsilon} \int \frac{dm^2}{(m^2)^{1+\epsilon}} \Theta_{\rm alg},
\label{eq-J2}
\end{eqnarray}
where the subscript $``q\to q(g)"$ represents the situation that only the quark $q$ is inside the jet while the gluon $g$ is radiated outside the jet. Here $m$ is again the invariant mass for the final-state parton pair, i.e. $q+g$. Since now only $q$ is inside the jet while the gluon $g$ is outside, $m$ is different from the jet invariant mass. 

In this case, it is the single quark that forms the jet, rather than quark and gluon together forms a jet. Following the algorithm constraint, we should have 
\bea
{\rm max}\left(J_{E_T}^{(n)}(q), J_{E_T}^{(n)}(g)\right) \geq J_{E_T}^{(n)}(q+g).
\label{e.q(g)}
\eea
This can be understood as follows. If $J_{E_T}^{(n)}(g)$ is the larger one among $\left(J_{E_T}^{(n)}(q), J_{E_T}^{(n)}(g)\right)$, and satisfies Eq.~\eqref{e.q(g)}, then we would have a single gluon form the jet. Since the jet algorithm is iterative, the gluon jet will be removed from the event. As a result, we are left with only a single quark, which will automatically form another jet. On the contrary, if $J_{E_T}^{(n)}(q)$ is the larger one among $\left(J_{E_T}^{(n)}(q), J_{E_T}^{(n)}(g)\right)$, the single quark first forms the jet. In other words, for a two-parton configuration, as long as the larger one among single-parton $J_{E_T}$ functions is greater than the $J_{E_T}$ function for the two-parton set, we will have the quark form the jet. With Eq.~\eqref{e.q(g)} at hand and from Eqs.~\eqref{eq:q+g} and \eqref{eq:qg}, we thus have the following constraint 
\bea
\Theta_{\rm alg} = \Theta \left[m^2 - \frac{E_{\mathrm{J}}^2}{z^2 \beta} {\rm min} \left[(1-z^n),(1-(1-z)^n)\right] \right]\,,
\label{eq-alg2}
\eea
where we have used $E=E_{\rm J}/z$ for this configuration. Implementing this constraint in Eq.~(\ref{eq-J2}), and considering the case $n=1$, we obtain
\begin{eqnarray}
J_{q\rightarrow q(g)}^{(I)}(z, E_{\mathrm{J}}) =\frac{\alpha_s}{2\pi} \frac{(\mu^2 e^{\gamma_E})^\epsilon}{\Gamma(1-\epsilon)} \left(\frac{1}{\epsilon}\right)  \left(\frac{E_{\mathrm{J}}^2}{\beta} \right)^{-\epsilon} \hat{P}_{qq}(z, \epsilon)
\left[(1-z)^{-2\epsilon }z^{\epsilon}\Theta(z-1/2)+(1-z)^{-\epsilon}\Theta(1/2-z) \right].
\end{eqnarray}
Performing the $\epsilon$-expansion, we find that the bare jet function is given as follows
\bea
J_{q \rightarrow q(g)}^{(I)}\left(z, E_{\mathrm{J}}\right)=& \frac{\alpha_{s}}{2 \pi} C_{F} \delta(1-z)\left(-\frac{1}{\epsilon^{2}}-\frac{1}{\epsilon} L-\frac{1}{2} L^{2}+\frac{\pi^{2}}{12}\right)
\nnu
&+\frac{\alpha_{s}}{2 \pi} C_{F}\left[\left(\frac{1}{\epsilon}+L\right) \frac{1+z^{2}}{(1-z)_{+}}-2\left(1+z^{2}\right)\left(\frac{\ln (1-z)}{1-z}\right)_{+}-(1-z)\right]
\nnu
&+\frac{\alpha_{s}}{2 \pi} C_{F}\frac{1+z^2}{1-z}\ln\left(\frac{1-z}{z}\right)\Theta(1/2-z), 
\label{eq-q2q}
\eea
where 
\bea
L=\hat{L}+\ln z=\ln \left(\frac{\beta \mu^2 z}{E_{\mathrm{J}}^2}\right).
\label{eq-L2}
\eea
We realize that Eq.~(\ref{eq-q2q}) is the same as that in the anti-$k_T$ algorithm, except for the slightly different definition of $L$. The universality of this term is caused by the NJA, and the algorithm dependence is hidden in power suppressed terms.

Likewise, for the case that only the gluon falls inside the jet, as illustrated in Fig.~\ref{fig-quark}(C), the calculation is very similar to the case of $q\to q(g)$.
For $n=1$, it can be expressed as
\bea
J_{q \rightarrow(q) g}^{(I)}\left(z, E_{\mathrm{J}}\right)=&\frac{\alpha_{s}}{2 \pi}\left(\frac{1}{\epsilon}+L\right) P_{g q}(z)-\frac{\alpha_{s}}{2 \pi} C_{F} \left[\frac{1+(1-z)^{2}}{z} 2 \ln (1-z)+ z\right]
\nnu
&+\frac{\alpha_{s}}{2 \pi} C_{F}\frac{1+(1-z)^{2}}{z}\ln\left(\frac{1-z}{z}\right)\Theta(1/2-z).
\label{eq-q2g}
\eea
Summing Eqs.~(\ref{eq-q2qg}, \ref{eq-q2q}, \ref{eq-q2g}) together, one obtains the full expression for the semi-inclusive quark jet function in the $J_{E_T}^{(I)}$ algorithm at NLO,
\bea
J_{q}^{(I)}\left(z, E_{\mathrm{J}}\right)=& J_{q \rightarrow q g}^{(I)}\left(z, E_{\mathrm{J}}\right)+J_{q \rightarrow q(g)}^{(I)} \left(z, E_{\mathrm{J}}\right)+J_{q \rightarrow(q) g}^{(I)} \left(z, E_{\mathrm{J}}\right) 
\nnu
=& \frac{\alpha_{s}}{2 \pi}\left(\frac{1}{\epsilon}+L\right)\left[P_{q q}(z)+P_{g q}(z)\right] 
-\frac{\alpha_{s}}{2 \pi}\Bigg\{C_F\left[2\left(1+z^{2}\right)\left(\frac{\ln (1-z)}{1-z}\right)_{+}+(1-z)\right]  
\nnu
& -\delta(1-z) C_F \left(\frac {3 \ln 2  }{2} + 5 - \frac{\pi ^2}{2}\right) +P_{gq}(z) 2 \ln (1-z)+C_{F} z 
\nnu
&-\left(P_{qq}(z)+P_{gq}(z)\right)\ln\left(\frac{1-z}{z}\right)\Theta(1/2-z)
\Bigg\},
\eea
where $P_{ij}(z)$ are the standard Altarelli-Parisi splitting functions,
\bea
P_{q q}(z)=&C_{F}\left[\frac{1+z^{2}}{(1-z)_{+}}+\frac{3}{2} \delta(1-z)\right],
\label{eq-Pqq}
\\
P_{g q}(z)=&C_{F} \frac{1+(1-z)^{2}}{z},
\label{eq-Pgq}
\\
P_{q g}(z)=&T_{F}\left(z^{2}+(1-z)^{2}\right),
\label{eq-Pqg}
\\
P_{g g}(z)=&C_{A}\left[\frac{2 z}{(1-z)_{+}}+\frac{2(1-z)}{z}+2 z(1-z)\right]+\frac{\beta_{0}}{2} \delta(1-z),
\label{eq-Pgg}
\eea 
with $\beta_{0}=\frac{11}{3} C_{A}-\frac{4}{3} T_F n_{f}$, and the ``plus"-distributions are defined as usual 
\bea
\int_{0}^{1} d z f(z)[g(z)]_{+} \equiv \int_{0}^{1} d z\left[f(z)-f(1)\right] g(z).
\eea
Note that the double pole term ($\propto 1/\epsilon^2$) as shown in $q\to qg$ channel cancel with that in $q\to q(g)$, and at the same time the $q\to (q)g$ channel is free of double-pole terms, thus we are left with only single-pole term (accordingly single logarithm $L$) in the final result. The single-pole term is universal and independent of jet algorithm, and it is this single-pole term that leads to the DGLAP evolution of the semi-inclusive jet functions~\cite{Kang:2016mcy}.

Likewise, the jet function for the $J_{E_T}^{(II)}$ algorithm with $n=2$ can be derived by following exactly the same procedure as that for $n=1$. The final result is
\bea
J_{q}^{(II)}\left(z, E_{\mathrm{J}}\right)=&\frac{\alpha_{s}}{2 \pi}\left(\frac{1}{\epsilon}+L\right)\left[{P}_{q q}(z)+P_{g q}(z)\right] 
\nnu
&-\frac{\alpha_{s}}{2 \pi}\Bigg\{C_F\left[2\left(1+z^{2}\right)\left(\frac{\ln (1-z)}{1-z}\right)_{+}+(1-z)\right] + P_{gq}(z) 2\ln (1-z) + C_{F} z
\nnu
&+ \left[P_{qq}(z)+P_{gq}(z)\right]\left[\ln (1+z)+\ln\left(\frac{z(2-z)}{1-z^2}\right)\Theta(1/2-z)\right] 
\nnu
& 
-\delta(1-z) C_F 
\left(-\frac{\pi^2}{2} + \frac{13}{2} - \frac{15\ln 2 }{2}- \ln^2 2  + \frac{9\ln 3}{2} - {\rm Li}_2(1/4)\right)\Bigg\}\,,
\eea
where ${\rm Li}_2(1/4)=0.267653\dots$ is the dilogarithm function. We find that all the terms in the last two rows are algorithm dependent, while the rest of the terms are independent of the jet algorithm. 

\subsection{The semi-inclusive gluon jet function}
The calculation of the semi-inclusive gluon jet function $J_g(z,E_{\mathrm{J}})$ is very similar to that for the quark jet case. As illustrated in Fig.~\ref{fig-gluon}, the semi-inclusive gluon jet function receives four contributions, including both $g\to gg$ and $g\to q\bar{q}$ splittings.
For the case that both final-state partons are inside the jet, the semi-inclusive gluon jet function can be written as
\begin{eqnarray}
J_{g\rightarrow gg + q\bar{q}}(z,E_{\mathrm{J}})= \delta(1-z) \frac{\alpha_{s}}{2 \pi}  \frac{(\mu^2 e^{\gamma_E})^\epsilon}{\Gamma(1-\epsilon)} \int_{0}^{1} dx \frac{1}{2} \left[2 n_f \hat P_{qg} (x,\epsilon)+ \hat{P}_{gg}(x,\epsilon)\right] \left[x(1-x)\right]^{-\epsilon} \int \frac{d m^2}{(m^2)^{1+\epsilon}} \Theta_{\rm alg},
\label{eq-gjet1}
\end{eqnarray}
where $n_f$ represents for the number of flavors of the final-state quark, and $\hat{P}_{qg}(x,\epsilon)$ and 
$\hat{P}_{gg}(x,\epsilon)$ are given by
Eqs.~(\ref{Pqghat}) and (\ref{Pgghat}), respectively.
In the situation when both the final state partons are inside the jet, the constraint is shown in Eq.~(\ref{eq-constraint2}). Apply this constraint to Eq.~(\ref{eq-gjet1}), we obtain the semi-inclusive gluon jet function 
\begin{eqnarray}
J_{g\rightarrow gg + q\bar{q}}^{(I)}(z,E_{\mathrm{J}}) = \frac{\alpha_{s}}{2 \pi}  \frac{(\mu^2 e^{\gamma_E})^\epsilon}{\Gamma(1-\epsilon)} \left(-\frac{1}{\epsilon}\right) \left(\frac{E_{\mathrm{J}}^2}{\beta}\right)^{-\epsilon} \frac{1}{2} \big[2 n_f I _{qg}  + I _{gg} \big],
\label{eq-g1}
\end{eqnarray}
where
\begin{eqnarray}
I_{ij} = \left[\int_{0}^{1 / 2} d x\left(1-(1-x)\right)^{-\epsilon}+\int_{1 / 2}^{1} d x\left(1-x\right)^{-\epsilon}\right] x^{-\epsilon}(1-x)^{-\epsilon} \hat P_{ij}(x, \epsilon).
\end{eqnarray}
Expanding to $\mathcal{O}(\epsilon)$, we obtain
\bea
I_{gg} =& 2 C_{A}\left[-\frac{1}{\epsilon}-\frac{11}{6}+\epsilon\left(-\frac{45}{8}+\frac{\pi^{2}}{2}-\frac{11}{6} \ln 2\right)\right],\\
I_{qg}=&T_F \left[\frac{2}{3}+\epsilon\left(\frac{23}{12}+\frac{2}{3} \ln 2\right)\right].
\eea
Insert the above expressions into Eq.~(\ref{eq-g1}) and perform the $\epsilon$-expansion, we find
\bea
J_{g \rightarrow gg+q\overline{q}}^{(I)}\left(z, E_{\mathrm{J}}\right)=&\delta(1-z) \frac{\alpha_{s}}{2 \pi}\Bigg[\frac{C_{A}}{\epsilon^{2}}+\frac{\beta_{0}}{2 \epsilon}+\frac{C_{A}}{\epsilon} \hat{L}+\frac{C_{A}}{2} \hat{L}^{2}+\frac{\beta_{0}}{2} \hat{L}
\nonumber\\
&+C_A \left(\frac{45}{8} + \frac{11}{6}\ln2 - \frac{7\pi^2}{12} \right) - n_f T_F \left(\frac{23}{12} + \frac{2}{3}\ln 2 \right)\Bigg]\,.
\eea
 
\bef
\includegraphics[width = 0.75\linewidth]{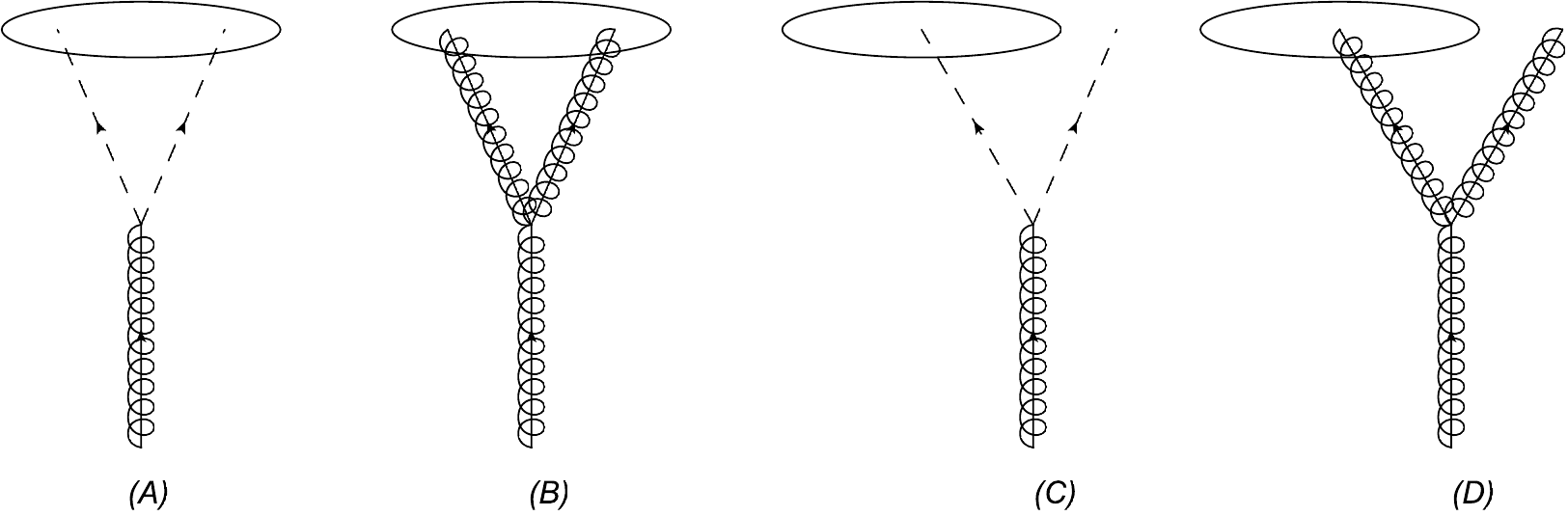}
\caption{The situations that contribute to the semi-inclusive gluon jet function when both final-state partons are inside the jet (A and B), and when only one of the final-state partons is inside the jet (C and D).}
\label{fig-gluon}	
\end{figure}

We also need to consider the situation that only one of the partons forms the jet and the corresponding Feynman diagrams are shown in Fig.~\ref{fig-gluon} (C) and (D). 
Summing these two diagrams together, we obtain 
\begin{eqnarray}
J_{g \rightarrow g(g)+q(\overline{q})}\left(z, E_{\mathrm{J}}\right)= \frac{\alpha_{s}}{2\pi} \frac{\left(\mu^{2} e^{\gamma_{E}}\right)^{\epsilon}}{\Gamma(1-\epsilon)}\left[\hat{P}_{g g}(z, \epsilon)+2 n_{f} \hat{P}_{q g}(z, \epsilon)\right]\left[z(1-z)\right]^{-\epsilon} \int \frac{d {m^2}}{(m^2)^{1+ \epsilon}} \Theta_{\mathrm{alg}}.
\end{eqnarray}
With the constraint from the $J_{E_T}^{(I)}$ algorithm shown in Eq.~(\ref{eq-alg2}), we can integrate over $m^2$ and then perform the $\epsilon$- expansion. The final results can be written as
\bea
J_{g \rightarrow g(g)+q(\overline{q})}^{(I)} \left(z, E_{\mathrm{J}}, \mu \right)=& \frac{\alpha_{s}}{2 \pi} \delta(1-z)\left(-\frac{C_{A}}{\epsilon^{2}}-\frac{\beta_{0}}{2 \epsilon}-\frac{C_{A}}{\epsilon} L-\frac{C_{A}}{2} L^{2}-\frac{\beta_{0}}{2} L+\frac{\pi^{2}}{12}\right) 
+\frac{\alpha_{s}}{2 \pi}\left(\frac{1}{\epsilon}+L\right)\left[P_{g g}(z)+2 n_{f} P_{q g}(z)\right] 
\nnu
&-\frac{\alpha_{s}}{2 \pi}\Bigg\{\frac{4 C_{A}\left(1-z+z^{2}\right)^{2}}{z}\left(\frac{\ln (1-z)}{1-z}\right)_{+} 
+4 n_{f}\left[P_{q g}(z) \ln (1-z)+T_{F} z(1-z)\right] 
\nnu
&+\left[P_{gg}(z)+2n_fP_{qg}(z)\right]\ln\left(\frac{1-z}{z}\right)\Theta(1/2-z)
\Bigg\}. 
\eea
Finally, adding all contributions together, we obtain the following expression for the semi-inclusive gluon jet function
\bea 
J_{g}^{(I)}\left(z, E_{\mathrm{J}} \right)=& J_{g \rightarrow g g+q \overline{q}}^{(I)}\left(z, E_{\mathrm{J}} \right)+J_{g \rightarrow g(g)+q(\overline{q})}^{(I)}\left(z, E_{\mathrm{J}} \right) 
\nnu
=& \frac{\alpha_{s}}{2 \pi}\left(\frac{1}{\epsilon}+L\right)\left[P_{g g}(z)+2 n_{f} P_{q g}(z)\right] 
-\frac{\alpha_{s}}{2 \pi} \Bigg \{\frac{4 C_{A}\left(1-z+z^{2}\right)^{2}}{z}\left(\frac{\ln (1-z)}{1-z}\right)_{+} 
+4 n_{f}\left[P_{q g}(z) \ln (1-z) \right.
 \nnu
& \left.+T_{F} z(1-z)\right] -\delta(1-z) \left[C_A \left(\frac{45}{8} + \frac{11}{6} \ln 2 - \frac{\pi^2}{2} \right) - n_f T_F \left(\frac{23}{12} + \frac{2}{3} \ln 2 \right)\right]  
\nnu
&-\left[P_{gg}(z)+2n_fP_{qg}(z)\right]\ln\left(\frac{1-z}{z}\right)\Theta(1/2-z)
\Bigg \},
\label{J1bare-final}
\eea
where $P_{g g}(z)$ and $P_{q g}(z)$ are the standard Altarelli-Parisi splitting functions defined in Eqs.~(\ref{eq-Pqg}) and (\ref{eq-Pgg}), respectively.
Similar to the case for the quark jet, the double pole $1/\epsilon^2$ and the double logarithms $L^2$ cancel, and we are left with only a single pole $1/\epsilon$ and a single logarithm $L$. 

Using the same procedure, we can also derive the semi-inclusive gluon jet function for $J_{E_T}^{(II)}$ algorithm. The final expression is given by
\bea
J_{g}^{(II)}\left(z, E_{\mathrm{J}} \right)=&\frac{\alpha_{s}}{2 \pi}\left(\frac{1}{\epsilon}+L\right)\left[{P}_{g g}(z)+2 n_{f} P_{q g}(z)\right] 
\nnu
&-\frac{\alpha_{s}}{2 \pi}\Bigg \{\frac{4 C_{A}\left(1-z+z^{2}\right)^{2}}{z}\left(\frac{\ln (1-z)}{1-z}\right)_{+}
+4 n_{f}\left[P_{q g}(z) \ln (1-z)+T_{F} z(1-z)\right]
\nnu
& -\left[ C_A\left(\frac{289}{36} - \frac{\pi^2}{2} -  \frac{79\ln 2}{6} - \ln^2 2 + \frac{15}{2}\ln 3 -  {\rm Li}_2 (1/4) \right) -n_f T_f \left(\frac{67}{18} - \frac{34\ln 2}{3} + 6 \ln 3 \right)\right] 
\nnu
& \times \delta (1-z)  + \left[P_{gg}(z)+2 n_{f}P_{qg}(z)\right] 
\left[\ln (1+z) + \ln\left(\frac{z(2-z)}{1-z^2}\right)\Theta(1/2-z) \right]
\Bigg\},
\label{J2bare-final}
\eea
where the last two rows in the above equation are algorithm dependent, while all the rest terms are algorithm independent.

\subsection{The RG evolution for semi-inclusive jet functions}
The bare quark and gluon jet functions $J_{q,g}(z, E_{\rm J})$ are divergent as they contain poles of $1/\epsilon$, which need to be renormalized. 
We follow the standard procedure and define the renormalized semi-inclusive jet functions $J_{q,g}(z, E_{\rm J}, \mu)$ as follows
\bea
J_{i}(z,E_{\mathrm{J}}) = \sum_{j} \int_{z}^{1} \frac{d z^\prime}{z^\prime} Z_{ij} \left(\frac{z}{z^\prime},\mu\right)J_j(z^\prime,E_{\mathrm{J}},\mu),
\eea
where $Z_{ij}$ is the renormalization matrix. Taking derivative with respect to $\mu$ on both sides of the above equation, we can obtain the renormalization-group equation for the semi-inclusive jet function 
\begin{equation}
\mu\frac{d}{d\mu}J_i(z,E_{\mathrm{J}},\mu) =  \sum_{j} \int_{z}^{1} \frac{d z^\prime}{z^\prime} \gamma_{ij}^J \left(\frac{z}{z^\prime},\mu\right)J_j(z^\prime,E_{\mathrm{J}},\mu),
\end{equation}
where $\gamma_{ij}^J$ is anomalous dimension related to the renormalization matrix
\begin{eqnarray}
\gamma_{ij}^J(z,\mu)=-\sum_{k} \int_{z}^{1}\frac{d z^\prime}{z^\prime} (Z)_{ik}^{-1} \left(\frac{z}{z^\prime},\mu\right) \mu\frac{d}{d\mu} Z_{kj}(z^\prime,\mu), 
\end{eqnarray}
with the inverse of the renormalization factor $(Z)_{ik}^{-1} $ defined as
\begin{eqnarray}
\sum_{k} \int_{z}^{1}\frac{d z^\prime}{z^\prime} (Z)_{ik}^{-1} \left(\frac{z}{z^\prime},\mu\right) Z_{kj}(z^\prime,\mu) = \delta_{ij}\delta(1-z).
\end{eqnarray}
At LO, the renormalization matrix $Z_{ij}^{(0)}$ is purly $\delta$-function, which is given by $Z_{ij}^{(0)}(z,\mu) = \delta_{ij}\delta(1-z)$. At NLO, the one-loop renormalization factors $Z_{ij}^{(1)}$ can be extracted from the pole terms in the final result of bare semi-inclusive jet function shown in Eqs.~(\ref{J1bare-final}) and (\ref{J2bare-final})
\begin{eqnarray}
Z_{ij}^{(1)}(z, \mu)=\frac{\alpha_s(\mu)}{2\pi}\left(\frac{1}{\epsilon}\right)P_{ji}(z)\,
\end{eqnarray}
from which we obtain the anomalous dimension 
\bea
\gamma_{ij}^J = \frac{\alpha_s(\mu)}{\pi} P_{ji}(z).
\eea
Thus the renormalization group equations for the renormalized siJFs is just the time-like DGLAP evolution equation for the usual fragmentation functions
\begin{equation}
\mu\frac{d}{d\mu}J_i(z,E_{\mathrm{J}},\mu) =  \frac{\alpha_s(\mu)}{\pi}\sum_{j} \int_{z}^{1} \frac{d z^\prime}{z^\prime} 
P_{ji}\left(\frac{z}{z^\prime}\right) J_j(z^\prime,E_{\mathrm{J}},\mu).
\label{eq-ren}
\end{equation}

The divergence of $Z_{ij}^{(1)}$ cancel exactly with the divergence in the bare semi-inclusive jet functions, and eventually leads to finite renormalized semi-inclusive jet functions. For the maximized algorithm with $n=1$, i.e. $J_{E_T}^{(I)}$, we have the following expressions for the renormalized siJFs at NLO
\bea
J_{q}^{(I)}\left(z, E_{\mathrm{J}},\mu\right)=& \delta(1-z) + \frac{\alpha_{s}}{2 \pi}L\left[P_{q q}(z)+P_{g q}(z)\right] 
-\frac{\alpha_{s}}{2 \pi}\Bigg\{C_F\left[2\left(1+z^{2}\right)\left(\frac{\ln (1-z)}{1-z}\right)_{+}+(1-z)\right]  
\nnu
& -\delta(1-z) C_F \left(\frac {3 \ln 2  }{2} + 5 - \frac{\pi ^2}{2}\right) +P_{gq}(z) 2 \ln (1-z)+C_{F} z 
\nnu
&-\left(P_{qq}(z)+P_{gq}(z)\right)\ln\left(\frac{1-z}{z}\right)\Theta(1/2-z)
\Bigg\},
\\
J_{g}^{(I)}\left(z, E_{\mathrm{J}},\mu \right) =& \delta(1-z) + \frac{\alpha_{s}}{2 \pi}L\left[P_{g g}(z)+2 n_{f} P_{q g}(z)\right] 
\nnu
&-\frac{\alpha_{s}}{2 \pi} \Bigg \{\frac{4 C_{A}\left(1-z+z^{2}\right)^{2}}{z}\left(\frac{\ln (1-z)}{1-z}\right)_{+} 
+4 n_{f}\left[P_{q g}(z) \ln (1-z)+T_{F} z(1-z)\right]
 \nnu
& -\delta(1-z) \left[C_A \left(\frac{45}{8} + \frac{11}{6} \ln 2 - \frac{\pi^2}{2} \right) - n_f T_F \left(\frac{23}{12} + \frac{2}{3} \ln 2 \right)\right] 
\nnu
&-\left[P_{gg}(z)+2n_fP_{qg}(z)\right]\ln\left(\frac{1-z}{z}\right)\Theta(1/2-z)
 \Bigg \}.
\eea
On the other hand, for the maximized algorithm with $n=2$, i.e. $J_{E_T}^{(II)}$, the siJFs at NLO can be written as follows
\bea 
J_{q}^{(II)}\left(z, E_{\mathrm{J}}, \mu\right)=&\delta(1-z) + \frac{\alpha_{s}}{2 \pi}L\left[{P}_{q q}(z)+P_{g q}(z)\right] 
\nnu
&-\frac{\alpha_{s}}{2 \pi}\Bigg\{C_F\left[2\left(1+z^{2}\right)\left(\frac{\ln (1-z)}{1-z}\right)_{+}+(1-z)\right] + P_{gq}(z) 2\ln (1-z) + C_{F} z
\nnu
&+ \left[P_{qq}(z)+P_{gq}(z)\right]\left[\ln (1+z) + \ln\left(\frac{z(2-z)}{1-z^2}\right)\Theta(1/2-z)\right] 
\nnu
& 
-\delta(1-z) C_F 
\left(-\frac{\pi^2}{2} + \frac{13}{2} - \frac{15\ln 2 }{2}- \ln^2 2  + \frac{9\ln 3}{2} - {\rm Li}_2(1/4)\right)\Bigg\}\,,
\\
J_{g}^{(II)}\left(z, E_{\mathrm{J}},\mu \right)=&\delta(1-z) + \frac{\alpha_{s}}{2 \pi}L\left[{P}_{g g}(z)+2 n_{f} P_{q g}(z)\right] 
\nnu
&-\frac{\alpha_{s}}{2 \pi}\Bigg \{\frac{4 C_{A}\left(1-z+z^{2}\right)^{2}}{z}\left(\frac{\ln (1-z)}{1-z}\right)_{+}
+4 n_{f}\left[P_{q g}(z) \ln (1-z)+T_{F} z(1-z)\right]
\nnu
& -\left[ C_A\left(\frac{289}{36} - \frac{\pi^2}{2} -  \frac{79\ln 2}{6} - \ln^2 2 + \frac{15}{2}\ln 3 -  {\rm Li}_2 (1/4) \right) -n_f T_f \left(\frac{67}{18} - \frac{34\ln 2}{3} + 6 \ln 3 \right)\right] 
\nnu
& \times \delta (1-z)  + \left[P_{gg}(z)+2 n_{f}P_{qg}(z)\right] 
\left[\ln (1+z) + \ln\left(\frac{z(2-z)}{1-z^2}\right)\Theta(1/2-z) \right]
\Bigg\}\,.
\eea

It is important to point out that the natural scale $\mu_{\rm J}$ for the siJFs can be derived from Eq.~\eqref{eq-L2},
\bea
\mu_{\rm J}^2 = \frac{E_{\rm J}^2}{\beta z}\,.
\eea
Thus by solving the DGLAP evolution equation Eq.~\eqref{eq-ren} and thus evolving the siJFs from its natural scale $\mu_{\rm J}$ to the hard scattering scale $\mu\sim E_{\rm J}$, we can resum large logarithms of the jet size parameter $\beta$. This is similar to the small jet radius $\ln R$ resummation developed in~\cite{Kang:2016mcy}. It might be worthwhile to point out that the momentum fraction $z$ also plays some role in this evolution, and it would be valuable to study numerically how important such an effect is. Finally, it is interesting to point out that the above renormalized semi-inclusive jet functions satisfy the following momentum sum rule~\cite{Dai:2016hzf, Kang:2017mda} 
\bea
\int_{0}^{1} dz\, z \, J_{i}(z, E_{\mathrm{J}}=zE ,\mu) = 1.
\eea 
This momentum conservation rule provides us an alternative way to check our results. Note that the above semi-inclusive jet functions were first derived in Ref.~\cite{Kaufmann:2015hma} via standard pQCD techniques~\footnote{Note that the above results are slightly different from those in the published version of Ref.~\cite{Kaufmann:2015hma}. However, those results are later updated, and we thank Werner Vogelsang for pointing this out.}.

\section{The semi-inclusive fragmenting jet functions }
\label{sec-fjf}
We now turn to the evaluation of jet substructure with maximized jet algorithms. In this section, we focus on the so-called semi-inclusive fragmenting jet functions (siFJFs) as initially introduced in Ref.~\cite{Kang:2016ehg}. The siFJFs describe the longitudinal momentum distribution of hadrons inside the jet. Comparing to the siJFs, we need one more variable, $\omega_h\approx 2E_h$,  to represent for the light-cone momentum of the observed hadron. Accordingly, $z_h = E_h/E_{\rm J}$ denotes the fraction of the jet energy carried by the observed hadron. In this section, we will first introduce the definition of siFJFs in SCET and we then present their NLO calculations for both the $J_{E_T}^{(I)}$ and $J_{E_T}^{(II)}$ algorithms. We will present the detailed steps of our calculations for the $J_{E_T}^{(I)}$ algorithm. At the end of the section, we will write down the results for the $J_{E_T}^{(II)}$ algorithm as the procedure is very similar. 

Similar to the siJFs, one can define the siFJFs for quark and gluon jets in terms of gauge invariant quark and gluon fields in SCET~\cite{Kang:2016ehg}
\bea
\mathcal{G}_{q}^{h}\left(z, z_{h}, E_{\mathrm{J}}\right)&=\frac{z}{2 N_{c}} \delta\left(z_{h}-\frac{E_{h}}{E_{\mathrm{J}}}\right) {\rm Tr}\left[\frac{\bar {\slashed{n}}}{2}\left\langle 0\left|\delta(\omega-\overline{n} \cdot \mathcal{P}) \chi_{n}(0)\right|(J h) X\right\rangle\left\langle(J h) X\left|\overline{\chi}_{n}(0)\right| 0\right\rangle\right],
\nnu
\mathcal{G}_{g}^{h}\left(z, z_{h}, E_{\mathrm{J}}\right)&=-\frac{z \,\omega}{(d-2)\left(N_{c}^{2}-1\right)} \delta\left(z_{h}-\frac{E_{h}}{E_{\mathrm{J}}}\right)\left\langle 0\left|\delta(\omega-\overline{n} \cdot \mathcal{P}) \mathcal{B}_{n \perp \mu}(0)\right|(J h) X\right\rangle \left\langle(J h) X\left|\mathcal{B}_{n \perp}^{\mu}(0)\right| 0\right\rangle,
\label{eq-Gdef}
\eea
where $(d-2)$ is the number of polarizations for initial gluons in $d$-dimension. The state $|(J h) X \rangle$ represents the final-state identified hadron ($h$) inside the  jet $J$ as denoted by $(Jh)$, $X$ stands for unobserved particles. On the other hand, $\omega$, $E_{\mathrm{J}}$ and $z$ have exactly the same meaning as those in siJFs. 

Different from the semi-inclusive jet functions as discussed in the previous section, the siFJFs involve hadronic scale, thus contains non-perturbative information and is in principle not calculable in pQCD. 
However, the matching coefficients between siFJFs and standard fragmentation functions can be determined perturbatively, thus one can simply replace the hadronic states in Eq.~(\ref{eq-Gdef}) by the corresponding partonic states, which allows us to use the methodology of pQCD. 
In the following, we will show explicitly the calculation of $\mathcal{G}_{i}^{j}$ with $i$ the initial parton and $j$ the fragmenting parton. Similar to siJFs, the siFJFs at LO involve only delta functions
\bea
\mathcal{G}_{i}^{j,(0)}\left(z, z_{h}, E_{\mathrm{J}}\right)=\delta_{ij} \delta(1-z) \delta\left(1-z_{h}\right).
\eea
The last two delta functions correspond to the case that the total energy of the initiating parton is transfered to the jet,  and the fragmenting parton carries the whole energy of the jet. 

\subsection{Fragmenting jet functions at NLO}
At NLO, the initiating parton splits into two partons, which leads to two contributions for siFJFs: one is for the case when both partons are inside the jet and another case is when only one parton is inside the jet. In the situation that both partons are inside the jet, the corresponding Feynman diagrams for quark and gluon initiated jet are shown in Figs. \ref{fig-quark}(A) and Fig.~\ref{fig-gluon}(A), respectively. In this case, all the initial parton energy $E$ stays inside the jet, thus $z=E_{\mathrm{J}}/E=1$. Different with the semi-inclusive jet functions, we now observe a particular fragmenting parton inside the jet, where the fragmenting parton carries only part of the jet energy as characterized by $z_h =E_{h}/E_{\mathrm{J}} < 1$. In $\overline{\rm MS}$ scheme, the one-loop bare siFJFs can be written as~\cite{Kang:2016ehg}
\bea
\mathcal{G}_{i, {\rm bare}}^{j k,(1)}\left(z, z_{h}, E_{\mathrm{J}}\right)=\delta(1-z) \frac{\alpha_{s}}{2\pi} \frac{\left(e^{\gamma_{E}} \mu^{2}\right)^{\epsilon}}{\Gamma(1-\epsilon)} \int_0^1 dx \, \delta(x - z_h) \hat{P}_{j i}\left(x, \epsilon\right)\, \left[x (1-x)\right]^{-\epsilon}\int \frac{d m^2}{(m^2)^{1+ \epsilon}} \Theta_{\mathrm{alg}}.
\label{eq-FJF_bare1}
\eea
In comparison with the siJFs in Eq.~\eqref{eq:siJFs}, the above expression is only different by a factor $\delta(x - z_h)$, since we now measure the momentum fraction carried by the final-state parton. Similar to that in semi-inclusive jet functions, $``jk"$ indicates the situation that both splitted partons $j$ and $k$ remain inside the jet. The functions $\hat{P}_{ji}(x,\epsilon)$ are given by Eqs.~(\ref{Pqqhat}),  (\ref{Pgqhat}), (\ref{Pqghat}), and (\ref{Pgghat}).

To proceed with the NLO calculation, we implement the jet algorithm constraint, given already in Eq.~\eqref{eq-constraint2}, and at the same time integrate $x$ with the help of the $\delta(x-z_h)$ function in Eq.~\eqref{eq-FJF_bare1}, we can derive the final result. It turns out that the result depends on the value of $z_h$. When $z_h \leq 1/2$, we have
\bea
\int \frac{d m^2}{(m^2)^{1+ \epsilon}} \Theta_{\mathrm{alg}}=\int_{0}^{\frac{E_{\mathrm{J}}^{2}}{\beta} (1-(1-z_h))}=\left(- \frac{1}{\epsilon} \right) \left(\frac{E_{\rm J}^2}{\beta}\right)^{-\epsilon} (1-(1-z_h))^{-\epsilon}.
\eea
While for $z_h \geq 1/2$, we have the following expression
\bea
\int \frac{d m^2}{(m^2)^{1+ \epsilon}} \Theta_{\mathrm{alg}}=\int_{0}^{\frac{E_{\rm J}^2}{\beta} (1-z_h)}=\left(- \frac{1}{\epsilon} \right) \left(\frac{E_{\rm J}^2}{\beta}\right)^{-\epsilon} (1-z_h)^{-\epsilon}.
\eea
Substituting these two integrals into Eq.~(\ref{eq-FJF_bare1}), we obtain the following result for the case when both partons are inside the jet 
\bea 
	\mathcal{G}_{i, {\rm bare}}^{j k,(1)} \left(z, z_{h}, E_{\mathrm{J}}\right)=&\delta(1-z) \frac{\alpha_{s}}{2 \pi} \frac{\left(e^{\gamma_{E}} \mu^{2}\right)^{\epsilon}}{\Gamma(1-\epsilon)} \hat{P}_{j i}\left(z_{h}, \epsilon\right)\left(-\frac{1}{\epsilon}\right) \left(\frac{E_{\rm J}^2}{\beta}\right)^{-\epsilon} \left[ (z_h)^{-2\epsilon} (1-z_h)^{-\epsilon}{\rm \theta}\left(\frac{1}{2}-z_h\right) \right.
	\nnu
&\left.	+ (z_h)^{-\epsilon} (1-z_h)^{-2\epsilon} {\rm \theta}\left(z_h-\frac{1}{2}\right) \right].
\label{eq-G1}
\eea

Now we turn to the case that only one parton is inside the jet. The corresponding diagrams are shown in Fig.~\ref{fig-quark} (B) and (C) for the quark-initiated jet and Fig.~\ref{fig-gluon} (C) and (D) for the gluon-initiated jet. In this case, the final-state fragmenting parton forms the jet, thus we have $z < 1$. On the other hand, within the jet, there is only one parton at NLO which is eventually converted to the fragmenting parton. This leads to an overall delta function ensuring $z_h=1$, i.e., all the jet energy is translated into the fragmenting parton energy. Performing the evaluation of diagrams Fig.~\ref{fig-quark} (B), (C) and Fig.~\ref{fig-gluon} (C), (D), we obtain the bare siFJFs when only parton $j$ forms the jet
\bea
\mathcal{G}_{i, {\rm bare}}^{j (k),(1)}\left(z, z_{h}, E_{\mathrm{J}}\right)=\delta(1-z_h) \frac{\alpha_{s}}{2\pi} \frac{\left(e^{\gamma_{E}} \mu^{2}\right)^{\epsilon}}{\Gamma(1-\epsilon)} \hat{P}_{j i}\left(z, \epsilon\right) z^{-\epsilon}(1-z)^{-\epsilon}\int \frac{d m^2}{(m^2)^{1+ \epsilon}} \Theta_{\mathrm{alg}},
\label{eq-FJF_bare2}
\eea
where the superscript ``$j(k)$'' indicates that parton $j$ is inside the jet while $k$ exits the jet. We can now implement the jet algorithm constraint Eq.~(\ref{eq-alg2}) in the $m^2$ integration, and we obtain for the $J_{E_T}^{(I)}$ algorithm
\bea
\int \frac{d m^2}{(m^2)^{1+ \epsilon}} \Theta_{\mathrm{alg}}
=\int_{\frac{E_{\rm J}^2}{z^2\beta} {\rm min} [1-z,z]}^{\infty} \frac{d m^2}{(m^2)^{1+ \epsilon}}  
=\left(\frac{1}{\epsilon} \right) \left(\frac{E_{\rm J}^2}{\beta}\right)^{-\epsilon} 
\left[(1-z)^{-\epsilon}z^{2\epsilon}\Theta(z-1/2) + z^{\epsilon}\Theta(1/2-z)\right].
\eea
With the above result at hand, we can write the second contribution of the bare siFJFs as follows
\bea
\mathcal{G}_{i, {\rm bare}}^{j(k),(1)}\left(z, z_{h}, E_{\mathrm{J}}\right)=\delta\left(1-z_{h}\right) \frac{\alpha_{s}}{2 \pi} \frac{\left(e^{\gamma_{E}} \mu^{2}\right)^{\epsilon}}{\Gamma(1-\epsilon)} \hat{P}_{j i}(z, \epsilon)\left(\frac{1}{\epsilon}\right)\left(\frac{E_{\rm J}^2}{\beta}\right)^{-\epsilon}
\left[(1-z)^{-2\epsilon}z^{\epsilon}\Theta(z-1/2) + (1-z)^{-\epsilon}\Theta(1/2-z)\right].
\label{eq-G2}
\eea
Summing the two contributions shown in Eqs.~(\ref{eq-G1}) and (\ref{eq-G2}), we obtain the following results for the bare siFJFs at NLO
\bea
\mathcal{G}_{i, {\rm bare}}^{j}\left(z, z_{h}, E_{\mathrm{J}}\right) = & \mathcal{G}_{i, {\rm bare}}^{j,(0)}\left(z, z_{h}, E_{\mathrm{J}}, \mu\right)+\mathcal{G}_{i, {\rm bare}}^{j k,(1)}\left(z, z_{h}, E_{\mathrm{J}}\right)+\mathcal{G}_{i, {\rm bare}}^{j(k),(1)}\left(z, z_{h}, E_{\mathrm{J}}\right)
\nnu
=&\delta_{ij} \delta(1-z) \delta(1-z_h) + \frac{\alpha_{s}}{2 \pi} \frac{\left(e^{\gamma_{E}} \mu^{2}\right)^{\epsilon}}{\Gamma(1-\epsilon)} \left(-\frac{1}{ \epsilon}\right) \left(\frac{E_{\rm J}^2}{\beta}\right)^{-\epsilon}
\nnu
& \times \Big\{\delta(1-z)\hat{P}_{j i}\left(z_{h}, \epsilon\right)\left[(z_h)^{-2\epsilon} (1-z_h)^{-\epsilon} \Theta(1/2-z_h)
+(z_h)^{-\epsilon} (1-z_h)^{-2\epsilon}\Theta(z_h-1/2)\right]
\nnu
&- \delta(1-z_h) \hat{P}_{j i}(z, \epsilon) \left[(1-z)^{-2\epsilon}z^{\epsilon}\Theta(z-1/2) + (1-z)^{-\epsilon}\Theta(1/2-z)\right]\Big\}.
\eea
Performing $\epsilon$-expansion, we obtain the explicit expression of the siFJFs for the $J_{E_T}^{(I)}$ algorithm
\bea
\mathcal{G}_{q, \mathrm{bare}}^{q,(I)}\left(z, z_{h}, E_{\mathrm{J}}\right)=&\delta(1-z) \delta\left(1-z_{h}\right)+\frac{\alpha_{s}}{2 \pi}\left(-\frac{1}{\epsilon}-L\right) P_{q q}\left(z_{h}\right) \delta(1-z) +\frac{\alpha_{s}}{2 \pi}\left(\frac{1}{\epsilon}+L\right) P_{q q}(z) \delta\left(1-z_{h}\right)
\nnu 
&+\delta(1-z) \frac{\alpha_{s}}{2 \pi}\left[2 C_{F}\left(1+z_{h}^{2}\right)\left(\frac{\ln \left(1-z_{h}\right)}{1-z_{h}}\right)_{+}+C_{F}\left(1-z_{h}\right)\right]\nnu
&+\delta(1-z) \frac{\alpha_{s}}{2 \pi} \left[ P_{qq}(z_h) \ln z_h + \Theta(1/2-z_h) P_{qq} (z_h) \ln \left(\frac{z_h}{1-z_h}\right)\right]
\nnu
&-\delta\left(1-z_{h}\right) \frac{\alpha_{s}}{2 \pi}\left[2 C_{F}\left(1+z^{2}\right)\left(\frac{\ln (1-z)}{1-z}\right)_{+}+C_{F}(1-z)
-P_{qq}(z)\ln\left(\frac{1-z}{z}\right)\Theta(1/2-z)\right],
\\
\mathcal{G}_{g, \mathrm{bare}}^{g,(I)}\left(z, z_{h}, E_{\mathrm{J}}\right)=&\delta(1-z) \delta\left(1-z_{h}\right)+\frac{\alpha_{s}}{2 \pi}\left(-\frac{1}{\epsilon}-L\right) P_{gg}\left(z_{h}\right) \delta(1-z) +\frac{\alpha_{s}}{2 \pi}\left(\frac{1}{\epsilon}+L\right) P_{gg}(z) \delta\left(1-z_{h}\right)
\nnu
&+\delta(1-z) \frac{\alpha_{s}}{2 \pi}\left[4 C_{A} \frac{\left(1-z_{h}+z_{h}^{2}\right)^{2}}{z_{h}}\left(\frac{\ln \left(1-z_{h}\right)}{1-z_{h}}\right)_{+}\right]
\nnu 
&+\delta(1-z) \frac{\alpha_{s}}{2 \pi} \left[ P_{gg}(z_h) \ln z_h + \Theta(1/2-z_h) P_{gg} (z_h) \ln \left(\frac{z_h}{1-z_h}\right)\right]
\nnu
&-\delta\left(1-z_{h}\right) \frac{\alpha_{s}}{2 \pi}\left[4 C_{A} \frac{\left(1-z+z^{2}\right)^{2}}{z}\left(\frac{\ln (1-z)}{1-z}\right)_{+}
-P_{gg}(z)\ln\left(\frac{1-z}{z}\right)\Theta(1/2-z)
\right],
\\
\mathcal{G}_{q, {\rm bare}}^{g,(I)}\left(z, z_{h}, E_{\mathrm{J}}\right)=&\frac{\alpha_{s}}{2 \pi}\left(-\frac{1}{\epsilon}-L\right) P_{g q}\left(z_{h}\right) \delta(1-z)+\frac{\alpha_{s}}{2 \pi}\left(\frac{1}{\epsilon}+L\right) P_{g q}(z) \delta\left(1-z_{h}\right)
\nnu
&+\delta(1-z) \frac{\alpha_{s}}{2 \pi}\left[2 P_{g q}\left(z_{h}\right) \ln \left(1-z_{h}\right)+C_{F} z_{h}\right]
\nnu
&+\delta(1-z) \frac{\alpha_{s}}{2 \pi} \left[ P_{gq}(z_h) \ln z_h + \Theta(1/2-z_h) P_{gq} (z_h) \ln \left(\frac{z_h}{1-z_h}\right)\right]
\nnu
&-\delta\left(1-z_{h}\right) \frac{\alpha_{s}}{2 \pi}\left[2 P_{g q}(z) \ln (1-z)+C_{F} z
-P_{gq}(z)\ln\left(\frac{1-z}{z}\right)\Theta(1/2-z)\right],
\\
\mathcal{G}_{g, \mathrm{bare}}^{q,(I)}\left(z, z_{h}, E_{\mathrm{J}}\right)=&\frac{\alpha_{s}}{2 \pi}\left(-\frac{1}{\epsilon}-L\right) P_{q g}\left(z_{h}\right) \delta(1-z)+\frac{\alpha_{s}}{2 \pi}\left(\frac{1}{\epsilon}+L\right) P_{q g}(z) \delta\left(1-z_{h}\right)
\nnu
&+\delta(1-z) \frac{\alpha_{s}}{2 \pi}\left[2 P_{q g}\left(z_{h}\right) \ln \left(1-z_{h}\right)+2 T_{F} z_{h}\left(1-z_{h}\right)\right]
\nnu
&+\delta(1-z) \frac{\alpha_{s}}{2 \pi} \left[P_{qg}(z_h) \ln z_h + \Theta(1/2-z_h) P_{qg} (z_h) \ln \left(\frac{z_h}{1-z_h}\right)\right]
\nnu
&-\delta\left(1-z_{h}\right) \frac{\alpha_{s}}{2 \pi}\left[2 P_{q g}(z) \ln (1-z)+2 T_{F} z(1-z)
-P_{qg}(z)\ln\left(\frac{1-z}{z}\right)\Theta(1/2-z)
\right],
\eea
where the logarithm $L$ is defined in Eq.~(\ref{eq-L2}), and the Altarelli-Parisi splitting functions are defined in Eqs.~(\ref{eq-Pqq}), (\ref{eq-Pgq}), (\ref{eq-Pqg}) and (\ref{eq-Pgg}).

The results for the bare siFJFs at NLO for the $J_{E_T}^{(II)}$ algorithm can be obtained similarly, and they are given by
\bea
\mathcal{G}_{q, \mathrm{bare}}^{q,(II)}\left(z, z_{h}, E_{\mathrm{J}}\right)=&\delta(1-z) \delta\left(1-z_{h}\right)+\frac{\alpha_{s}}{2 \pi}\left(-\frac{1}{\epsilon}-L\right) {P}_{qq}\left(z_{h}\right) \delta(1-z) +\frac{\alpha_{s}}{2 \pi}\left(\frac{1}{\epsilon}+L\right) {P}_{q q}(z) \delta\left(1-z_{h}\right)
\nnu
&+\delta(1-z) \frac{\alpha_{s}}{2 \pi}\left[2 C_{F}\left(1+z_{h}^{2}\right)\left(\frac{\ln \left(1-z_{h}\right)}{1-z_{h}}\right)_{+}+C_{F}\left(1-z_{h}\right)\right]
\nnu
&+\delta(1-z) \frac{\alpha_{s}}{2 \pi} \left[ \tilde{P}_{qq}(z_h) \ln \left(z_h(1+z_h)\right) + \Theta(1/2-z_h) {P}_{qq} (z_h) \ln \left(\frac{z_h(2-z_h)}{1-z_h^2}\right)\right]
\nnu
&-\delta\left(1-z_{h}\right) \frac{\alpha_{s}}{2 \pi}\Bigg[2 C_{F}\left(1+z^{2}\right)\left(\frac{\ln (1-z)}{1-z}\right)_{+}+C_{F}\left(1+z^{2}\right)\frac{\ln (1+z)}{(1-z)_+}+C_{F}(1-z)
\nnu
&+P_{qq}(z)\ln\left(\frac{z(2-z)}{1-z^2}\right)\Theta(1/2-z)
\Bigg],
\\
\mathcal{G}_{g, \mathrm{bare}}^{g,(II)}\left(z, z_{h}, E_{\mathrm{J}}\right)=&\delta(1-z) \delta\left(1-z_{h}\right)+\frac{\alpha_{s}}{2 \pi}\left(-\frac{1}{\epsilon}-L\right) {P}_{gg}\left(z_{h}\right) \delta(1-z) +\frac{\alpha_{s}}{2 \pi}\left(\frac{1}{\epsilon}+L\right) {P}_{gg}(z) \delta\left(1-z_{h}\right)
\nnu
&+\delta(1-z) \frac{\alpha_{s}}{2 \pi}\left[4 C_{A} \frac{\left(1-z_{h}+z_{h}^{2}\right)^{2}}{z_{h}}\left(\frac{\ln \left(1-z_{h}\right)}{1-z_{h}}\right)_{+}\right]
\nnu
&+\delta(1-z) \frac{\alpha_{s}}{2 \pi} \left[ \tilde{P}_{gg}(z_h) \ln \left(z_h(1+z_h)\right) + \Theta(1/2-z_h) {P}_{gg} (z_h) \ln \left(\frac{z_h(2-z_h)}{1-z_h^2}\right)\right]
\nnu
&-\delta\left(1-z_{h}\right) \frac{\alpha_{s}}{2 \pi}\Bigg[4 C_{A} \frac{\left(1-z+z^{2}\right)^{2}}{z}\left(\frac{\ln (1-z)}{1-z}\right)_{+}+2 C_{A} \frac{\left(1-z+z^{2}\right)^{2}}{z}\frac{\ln (1+z)}{(1-z)_+}
\nnu
&+P_{gg}(z)\ln\left(\frac{z(2-z)}{1-z^2}\right)\Theta(1/2-z)
\Bigg],
\\
\mathcal{G}_{q, {\rm bare}}^{g,(II)}\left(z, z_{h}, E_{\mathrm{J}}\right)=&\frac{\alpha_{s}}{2 \pi}\left(-\frac{1}{\epsilon}-L\right) P_{g q}\left(z_{h}\right) \delta(1-z)+\frac{\alpha_{s}}{2 \pi}\left(\frac{1}{\epsilon}+L\right) P_{g q}(z) \delta\left(1-z_{h}\right)
\nnu
&+\delta(1-z) \frac{\alpha_{s}}{2 \pi}\left[2 P_{g q}\left(z_{h}\right) \ln \left(1-z_{h}\right)+C_{F} z_{h}\right]
\nnu
&+\delta(1-z) \frac{\alpha_{s}}{2 \pi} \left[P_{gq}(z_h) \ln \left(z_h(1+z_h)\right) + \Theta(1/2-z_h) P_{gq} (z_h) \ln \left(\frac{z_h(2-z_h)}{1-z_h^2}\right)\right]
\nnu
&-\delta\left(1-z_{h}\right) \frac{\alpha_{s}}{2 \pi}\Bigg[2 P_{g q}(z) \ln (1-z)+P_{g q}(z) \ln (1+z)+C_{F} z
\nnu
&+P_{gq}(z)\ln\left(\frac{z(2-z)}{1-z^2}\right)\Theta(1/2-z)
\Bigg],
\\
\mathcal{G}_{g, \mathrm{bare}}^{q,(II)}\left(z, z_{h}, E_{\mathrm{J}}\right)=&\frac{\alpha_{s}}{2 \pi}\left(-\frac{1}{\epsilon}-L\right) P_{q g}\left(z_{h}\right) \delta(1-z)+\frac{\alpha_{s}}{2 \pi}\left(\frac{1}{\epsilon}+L\right) P_{q g}(z) \delta\left(1-z_{h}\right)
\nnu
&+\delta(1-z) \frac{\alpha_{s}}{2 \pi}\left[2 P_{q g}\left(z_{h}\right) \ln \left(1-z_{h}\right)+2 T_{F} z_{h}\left(1-z_{h}\right)\right]
\nnu
&+\delta(1-z) \frac{\alpha_{s}}{2 \pi} \left[P_{qg}(z_h) \ln \left(z_h(1+z_h)\right) + \Theta(1/2-z_h) P_{qg} (z_h) \ln \left(\frac{z_h(2-z_h)}{1-z_h^2}\right)\right]
\nnu
&-\delta\left(1-z_{h}\right) \frac{\alpha_{s}}{2 \pi}\Bigg[2 P_{q g}(z) \ln (1-z)+P_{q g}(z) \ln (1+z)+2 T_{F} z(1-z)
\nnu
& + P_{qg}(z)\ln\left(\frac{z(2-z)}{1-z^2}\right)\Theta(1/2-z)\Bigg].
\eea
where $\tilde P_{qq}(z)$ and $\tilde P_{gg}(z)$ are the Altarelli-Parisi splitting functions defined in Eqs.~(\ref{eq-Pqq}) and (\ref{eq-Pgg}), but with the $\delta(1-z)$-term removed. For later convenience, we also define $\tilde P_{gq}(z) = P_{gq}(z)$ and $\tilde P_{qg}(z) = P_{qg}(z)$. Notice that the pole terms are universal for different jet algorithms. This is why all these semi-inclusive jet functions follow the same time-like DGLAP evolution equations~\cite{Kang:2016ehg}.
 
\subsection{Renormalization and matching}
From the final expressions of the bare siFJFs $\mathcal{G}_{i}^{j}\left(z, z_{h}, E_{\mathrm{J}}\right)$, single pole terms still remain, which include both infrared (IR) poles as identified in association with $P_{ji}(z_h)\delta(1-z)$, and the ultraviolet (UV) poles that multiplied by $P_{ji}(z)\delta(1-z_h)$. The standard procedure to subtract the UV poles is realized by the definition of renormalized siFJFs. After the renormalization, we find the following renormalization group equations for the renormalized siFJFs $\mathcal{G}_{i}^{h}\left(z, z_{h}, E_{\mathrm{J}}, \mu\right)$ 
\bea
 \mu\frac{d}{d\mu}\mathcal{G}_{i}^{h}\left(z, z_{h}, E_{\mathrm{J}}, \mu\right) = \frac{\alpha_s(\mu)}{\pi}  \sum_j \int_z^1\frac{dz'}{z'}
P_{ji}\left(\frac{z}{z'}\right)\mathcal{G}_{k}^{j}\left(z', z_{h}, E_{\mathrm{J}}, \mu\right),
\eea
which is the same as the usual time-like DGLAP evolution equation, just like that for the siJFs in Eq.~\eqref{eq-ren}. From the perturbative calculations of $\mathcal{G}_{i}^{h}\left(z, z_{h}, E_{\mathrm{J}}, \mu\right)$, we find that the natural scale for the siFJFs is again $\mu_{\mathcal G}^2\sim E_J^2/\beta z$. Thus solving the above evolution equations and evolving the siFJFs from $\mu_{\mathcal G}$ to the typical hard scale $\mu\sim E_J$, one again achieves the resummation of $\ln\beta$. This could be very important for highly collimated jets where $\beta \gg 1$. 

It is important to emphasize that the renormalization equations for siFJFs are universal and independent of specific jet algorithms. Different to the siJFs shown in Sec.~\ref{sec-jet}, there are explicit flavor dependence for siFJFs due to the involved fragmentation functions. This leads to more complicated flavor separation in solving the renormalization equations for siFJFs than that for siJFs. For details, see Ref.~\cite{Kang:2016ehg}. After renormalization, the remaining IR poles are removed by matching the renormalized siFJFs $\mathcal{G}_{i}^{j}\left(z, z_{h}, E_{\mathrm{J}}, \mu\right) $ onto the fragmentation functions $D_{i}^{h}(z_h, \mu)$ at a scale $\mu \gg \Lambda_{\mathrm{QCD}}$ as follows
 \bea
  \mathcal{G}_{i}^{h}\left(z, z_{h}, E_{\mathrm{J}}, \mu\right)
  =\sum_{j} \int_{z_{h}}^{1} 
  \frac{d z_h^\prime}{z_h^\prime} \mathcal{J}_{i j}\left(z, z_{h}^{\prime}, E_{\rm J}, \mu\right) D_{j}^{h}\left(\frac{z_h}{z_h^\prime}, \mu\right).
  \eea
  
To derive the matching coefficients $\mathcal{J}_{i j}$, we again replace the hadron $h$ by a parton state in the above equation. Note that the renormalized fragmentation functions $D_{i}^{j}\left(z_{h}, \mu\right)$ at NLO in the 
$\overline{\mathrm{MS}}$ scheme are given by
\bea
 D_{i}^{j}\left(z_{h}, \mu\right)=\delta_{i j} \delta\left(1-z_{h}\right)+\frac{\alpha_{s}}{2 \pi} P_{j i}\left(z_{h}\right)\left(-\frac{1}{\epsilon}\right)\,.
\eea
With these results at hand, and using the expressions for $\mathcal{G}_{i}^{j}\left(z, z_{h}, E_{\mathrm{J}}, \mu\right)$ from the previous section, one can obtain the matching coefficient functions $\mathcal{J}_{i j}$, which are free of any divergences. For the maximized jet algorithm, they can be expressed as follows
\bea
  \mathcal{J}_{qq}\left(z, z_{h}, E_{\mathrm{J}}, \mu\right)=&\delta(1-z) \delta\left(1-z_{h}\right)+\frac{\alpha_{s}}{2 \pi}\Bigg\{L \left[P_{q q}\left(z\right) \delta(1-z_h) -  P_{q q}(z_h) \delta\left(1-z\right)\right]
  \nnu
  &+\delta(1-z) \left[2 C_{F}\left(1+z_{h}^{2}\right)\left(\frac{\ln \left(1-z_{h}\right)}{1-z_{h}}\right)_{+}+C_{F}\left(1-z_{h}\right)+\mathcal{I}_{q q}^{\rm alg}\left(z_{h}\right)\right]
  \nnu
  &\left.-\delta\left(1-z_{h}\right) \left[2 C_{F}\left(1+z^{2}\right)\left(\frac{\ln (1-z)}{1-z}\right)_{+}+C_{F}(1-z)+\mathcal{I}_{q q}^{'\rm alg}(z)\right]
  \right\},
  \\
  \mathcal{J}_{gg}\left(z, z_{h}, E_{\mathrm{J}}, \mu\right)=&\delta(1-z) \delta\left(1-z_{h}\right)+\frac{\alpha_{s}}{2 \pi}\left\{L\left[ P_{gg}\left(z\right) \delta(1-z_h) - P_{gg}(z_h) \delta(1-z)\right]\right.
  \nnu
  &+\delta(1-z) \left[4 C_{A} \frac{\left(1-z_{h}+z_{h}^{2}\right)^{2}}{z_{h}}\left(\frac{\ln \left(1-z_{h}\right)}{1-z_{h}}\right)_{+}+\mathcal{I}_{gg}^{\rm alg}\left(z_{h}\right)\right]
  \nnu
  &\left.-\delta\left(1-z_{h}\right) \left[4 C_{A} \frac{\left(1-z+z^{2}\right)^{2}}{z}\left(\frac{\ln (1-z)}{1-z}\right)_{+}
  +\mathcal{I}_{gg}^{'\rm alg}(z)\right]\right\},
  \\
  \mathcal{J}_{qg}\left(z, z_{h}, E_{\mathrm{J}}, \mu\right)=&\frac{\alpha_{s}}{2 \pi}\Big\{ L \left[P_{g q}\left(z\right) \delta(1-z_h)- P_{g q}(z_h) \delta\left(1-z\right)\right]
  \nnu
  &+\delta(1-z) \left[2 P_{g q}\left(z_{h}\right) \ln \left(1-z_{h}\right)+C_{F} z_{h}+ \mathcal{I}_{q g}^{\rm alg}\left(z_{h}\right)\right]
  \nnu
  &-\delta\left(1-z_{h}\right) \left[2 P_{g q}(z) \ln (1-z)+C_{F} z + \mathcal{I}_{qg}^{'\rm alg}(z)\right]\Big\},
  \\
  \mathcal{J}_{gq}\left(z, z_{h}, E_{\mathrm{J}}, \mu\right)=&\frac{\alpha_{s}}{2 \pi}\Big\{L\left[P_{q g}\left(z\right) \delta(1-z_h)- P_{q g}(z_h) \delta\left(1-z\right)\right]
  \nnu
  &+\delta(1-z) \left[2 P_{q g}\left(z_{h}\right) \ln \left(1-z_{h}\right)+2 T_{F} z_{h}\left(1-z_{h}\right)+\mathcal{I}_{gq}^{\rm alg}\left(z_{h}\right)\right]
  \nnu
  &-\delta\left(1-z_{h}\right) \left[2 P_{q g}(z) \ln (1-z)+2 T_{F} z(1-z)
  +\mathcal{I}_{gq}^{'\rm alg}(z)\right]\Big\},
  \eea
where $\mathcal{I}_{ij}^{\rm alg}(z_{h})$ and $\mathcal{I}_{ij}^{'\rm alg}(z_{h})$ are algorithm-dependent functions with the following expressions for the $J_{E_T}^{(I)}$ and $J_{E_T}^{(II)}$ algorithms
  \bea
  \mathcal{I}_{ij}^{(I)}(z_{h}) =& P_{ji}(z_h) \ln z_h + \Theta(1/2-z_h) P_{ji} (z_h) \ln \left(\frac{z_h}{1-z_h}\right),
  \\
   \mathcal{I}_{ij}^{(II)}(z_{h})= & \tilde{P}_{ji}(z_h) \ln \left(z_h(1+z_h)\right) + \Theta(1/2-z_h) {P}_{ji} (z_h) \ln \left(\frac{z_h(2-z_h)}{1-z_h^2}\right),
   \\
   \mathcal{I}_{ij}^{'(I)}(z) =&  -\Theta(1/2-z) P_{ji} (z) \ln \left(\frac{1-z}{z}\right),
  \\
   \mathcal{I}_{ij}^{'(II)}(z)= &  \tilde{P}_{ji}(z)\ln(1+z) + \Theta(1/2-z) {P}_{ji} (z) \ln \left(\frac{z(2-z)}{1-z^2}\right).
  \eea
 
Similar to the standard collinear fragmentation functions, the siFJFs satisfy the momentum sum rules~\cite{Procura:2009vm,Procura:2011aq}. These can be represented by the relations between the siJFs and the siFJFs 
  \bea
  \int_{0}^{1} dz_h z_h [\mathcal{G}_{q}^{q} (z,z_h,E_{\mathrm{J}},\mu) +\mathcal{G}_{q}^{g} (z,z_h,E_{\mathrm{J}},\mu)] &=  J_q(z,E_{\mathrm{J}},\mu),
   \\
  \int_{0}^{1} dz_h z_h [\mathcal{G}_{g}^{g} (z,z_h,E_{\mathrm{J}},\mu) +2 n_f\,\mathcal{G}_{g}^{q} (z,z_h,E_{\mathrm{J}},\mu)] &=  J_g(z,E_{\mathrm{J}},\mu).
  \eea
Or equivalently, we have
  \bea
   \int_{0}^{1} d z_{h} z_{h}\left[\mathcal{J}_{q q}\left(z, z_{h}, E_{\mathrm{J}}, \mu\right)+\mathcal{J}_{q g}\left(z, z_{h}, E_{\mathrm{J}}, \mu\right)\right] &=J_{q}\left(z, E_{\mathrm{J}}, \mu\right),
    \\ 
   \int_{0}^{1} d z_{h} z_{h}\left[\mathcal{J}_{g g}\left(z, z_{h}, E_{\mathrm{J}}, \mu\right)+2 n_{f}\, \mathcal{J}_{g q}\left(z, z_{h}, E_{\mathrm{J}}, \mu\right)\right] &=J_{g}\left(z, E_{\mathrm{J}}, \mu\right). 
  \eea
 The above two equations provide us a good way to double check our final results. We have checked for both $J_{E_T}^{(I)}$ and $J_{E_T}^{(II)}$ algorithms, our results satisfy the above momentum sum rules. Note that the result for the $J_{E_T}^{(I)}$ algorithm has first been derived from the standard pQCD method in~\cite{Kaufmann:2015hma}~\footnote{Again, our results are slightly different from those in \cite{Kaufmann:2015hma}, which are later updated.}. On the other hand, the results of the siFJFs for the $J_{E_T}^{(II)}$ algorithm are written down here for the first time as far as we know.

\section{conclusion}
\label{sec-sum}
In this paper, we evaluated the semi-inclusive jet functions (siJFs) and semi-inclusive fragmenting jet functions (siFJFs) at next-to-leading order (NLO), for both quark and gluon jets in the so-called maximized jet algorithms $J_{E_T}^{(I)}$ and $J_{E_T}^{(II)}$ within the framework of soft-collinear effective theory. Our fixed order results are consistent with the NLO results using standard pQCD techniques when available, and our result for semi-inclusive fragmenting jet functions for the $J_{E_T}^{(II)}$ algorithm are new. We further derived the renormalization group equations for both the siJFs and siFJFs, and found that they follow the time-like DGLAP evolution equations, just like the usual collinear fragmentation functions. These maximized jet algorithms contain a parameter $\beta$, which controls the geometric size of the reconstructed jet. $\beta \gg 1$ corresponds to a highly collimated jet, very similar to the situation for jets with a very small jet radius $R$. By solving the renormalization equations and evolving those jet functions from their natural scale to the typical hard scale, one can then achieve the resummation of the single-logarithms of the jet parameter $\beta$. In addition, we find that the evolution equations can further resum $\ln z$, where $z$ is the momentum fraction of the parton that initiates the jet carried by the jet itself. Phenomenological implementations of our results to hadronic collisions will be left for future publications. We expect important impact of our results in probing fundamental properties of nuclear medium and hadron fragmentation functions, as pointed out already from other fixed-order calculations, see for example, in Ref.~\cite{Kaufmann:2015hma}.  

\section*{Acknowledgments}
We thank Werner Vogelsang for very useful comments and communications.  L.W. and B.Z. are supported by the NSFC of China under Project No.~11805167, Z.K. is supported by the National Science Foundation in US under Grant No. PHY-1720486, and H.X. is supported by NSFC of China under Project No.~11435004 and research startup funding at SCNU.


\begin{thebibliography}{10}

\bibitem{Sterman:1977wj}
G.~F. Sterman and S.~Weinberg,
\newblock Phys. Rev. Lett. {\bf 39}, 1436 (1977).

\bibitem{Ellis:2007ib}
S.~D. Ellis, J.~Huston, K.~Hatakeyama, P.~Loch, and M.~Tonnesmann,
\newblock Prog. Part. Nucl. Phys. {\bf 60}, 484 (2008), arXiv:0712.2447.

\bibitem{Sapeta:2015gee}
S.~Sapeta,
\newblock Prog. Part. Nucl. Phys. {\bf 89}, 1 (2016), arXiv:1511.09336.

\bibitem{Buttar:2008jx}
C.~Buttar {\em et~al.},
\newblock {Standard Model Handles and Candles Working Group: Tools and Jets
  Summary Report},
\newblock in {\em {Physics at TeV colliders, La physique du TeV aux
  collisionneurs, Les Houches 2007 : 11-29 June 2007}}, pp. 121--214, 2008,
  arXiv:0803.0678.

\bibitem{Salam:2009jx}
G.~P. Salam,
\newblock Eur. Phys. J. {\bf C67}, 637 (2010), arXiv:0906.1833.

\bibitem{Altheimer:2012mn}
A.~Altheimer {\em et~al.},
\newblock J. Phys. {\bf G39}, 063001 (2012), arXiv:1201.0008.

\bibitem{Lai:1996mg}
H.~L. Lai {\em et~al.},
\newblock Phys. Rev. {\bf D55}, 1280 (1997), arXiv:hep-ph/9606399.

\bibitem{Martin:2001es}
A.~D. Martin, R.~G. Roberts, W.~J. Stirling, and R.~S. Thorne,
\newblock Eur. Phys. J. {\bf C23}, 73 (2002), arXiv:hep-ph/0110215.

\bibitem{Aschenauer:2019uex}
E.-C. Aschenauer, K.~Lee, B.~S. Page, and F.~Ringer,
\newblock (2019), arXiv:1910.11460.

\bibitem{Arratia:2019vju}
M.~Arratia, Y.~Song, F.~Ringer, and B.~Jacak,
\newblock (2019), arXiv:1912.05931.

\bibitem{Stump:2003yu}
D.~Stump {\em et~al.},
\newblock JHEP {\bf 10}, 046 (2003), arXiv:hep-ph/0303013.

\bibitem{Butterworth:2008iy}
J.~M. Butterworth, A.~R. Davison, M.~Rubin, and G.~P. Salam,
\newblock Phys. Rev. Lett. {\bf 100}, 242001 (2008), arXiv:0802.2470.

\bibitem{Vitev:2008rz}
I.~Vitev, S.~Wicks, and B.-W. Zhang,
\newblock JHEP {\bf 11}, 093 (2008), arXiv:0810.2807.

\bibitem{Vitev:2009rd}
I.~Vitev and B.-W. Zhang,
\newblock Phys. Rev. Lett. {\bf 104}, 132001 (2010), arXiv:0910.1090.

\bibitem{Muller:2012zq}
B.~Muller, J.~Schukraft, and B.~Wyslouch,
\newblock Ann. Rev. Nucl. Part. Sci. {\bf 62}, 361 (2012), arXiv:1202.3233.

\bibitem{Armesto:2015ioy}
N.~Armesto and E.~Scomparin,
\newblock Eur. Phys. J. Plus {\bf 131}, 52 (2016), arXiv:1511.02151.

\bibitem{Connors:2017ptx}
M.~Connors, C.~Nattrass, R.~Reed, and S.~Salur,
\newblock Rev. Mod. Phys. {\bf 90}, 025005 (2018), arXiv:1705.01974.

\bibitem{Larkoski:2017jix}
A.~J. Larkoski, I.~Moult, and B.~Nachman,
\newblock Phys. Rept. {\bf 841}, 1 (2020), arXiv:1709.04464.

\bibitem{Wang:2016opj}
X.-N. Wang, editor,
\newblock {\em {Quark-Gluon Plasma 5}} (World Scientific, New Jersey, 2016).

\bibitem{Page:2019gbf}
B.~S. Page, X.~Chu, and E.~C. Aschenauer,
\newblock (2019), arXiv:1911.00657.

\bibitem{Bauer:2000ew}
C.~W. Bauer, S.~Fleming, and M.~E. Luke,
\newblock Phys. Rev. {\bf D63}, 014006 (2000), arXiv:hep-ph/0005275.

\bibitem{Bauer:2000yr}
C.~W. Bauer, S.~Fleming, D.~Pirjol, and I.~W. Stewart,
\newblock Phys. Rev. {\bf D63}, 114020 (2001), arXiv:hep-ph/0011336.

\bibitem{Bauer:2001ct}
C.~W. Bauer and I.~W. Stewart,
\newblock Phys. Lett. {\bf B516}, 134 (2001), arXiv:hep-ph/0107001.

\bibitem{Bauer:2001yt}
C.~W. Bauer, D.~Pirjol, and I.~W. Stewart,
\newblock Phys. Rev. {\bf D65}, 054022 (2002), arXiv:hep-ph/0109045.

\bibitem{Kang:2016mcy}
Z.-B. Kang, F.~Ringer, and I.~Vitev,
\newblock JHEP {\bf 10}, 125 (2016), arXiv:1606.06732.

\bibitem{Dai:2016hzf}
L.~Dai, C.~Kim, and A.~K. Leibovich,
\newblock Phys. Rev. {\bf D94}, 114023 (2016), arXiv:1606.07411.

\bibitem{Kaufmann:2015hma}
T.~Kaufmann, A.~Mukherjee, and W.~Vogelsang,
\newblock Phys. Rev. {\bf D92}, 054015 (2015), arXiv:1506.01415.

\bibitem{Kang:2016ehg}
Z.-B. Kang, F.~Ringer, and I.~Vitev,
\newblock JHEP {\bf 11}, 155 (2016), arXiv:1606.07063.

\bibitem{Kang:2017frl}
Z.-B. Kang, F.~Ringer, and I.~Vitev,
\newblock Phys. Lett. {\bf B769}, 242 (2017), arXiv:1701.05839.

\bibitem{Fickinger:2016rfd}
M.~Fickinger, S.~Fleming, C.~Kim, and E.~Mereghetti,
\newblock JHEP {\bf 11}, 095 (2016), arXiv:1606.07737.

\bibitem{Dai:2018ywt}
L.~Dai, C.~Kim, and A.~K. Leibovich,
\newblock JHEP {\bf 09}, 109 (2018), arXiv:1805.06014.

\bibitem{Li:2018xuv}
H.~T. Li and I.~Vitev,
\newblock JHEP {\bf 07}, 148 (2019), arXiv:1811.07905.

\bibitem{Kang:2017yde}
Z.-B. Kang, J.-W. Qiu, F.~Ringer, H.~Xing, and H.~Zhang,
\newblock Phys. Rev. Lett. {\bf 119}, 032001 (2017), arXiv:1702.03287.

\bibitem{Bain:2017wvk}
R.~Bain, L.~Dai, A.~Leibovich, Y.~Makris, and T.~Mehen,
\newblock Phys. Rev. Lett. {\bf 119}, 032002 (2017), arXiv:1702.05525.

\bibitem{Dasgupta:2014yra}
M.~Dasgupta, F.~Dreyer, G.~P. Salam, and G.~Soyez,
\newblock JHEP {\bf 04}, 039 (2015), arXiv:1411.5182.

\bibitem{Georgi:2014zwa}
H.~Georgi,
\newblock (2014), arXiv:1408.1161.

\bibitem{Bai:2014qca}
Y.~Bai, Z.~Han, and R.~Lu,
\newblock JHEP {\bf 03}, 102 (2015), arXiv:1411.3705.

\bibitem{Kaufmann:2014nda}
T.~Kaufmann, A.~Mukherjee, and W.~Vogelsang,
\newblock Phys. Rev. {\bf D91}, 034001 (2015), arXiv:1412.0298.

\bibitem{Ellis:2010rwa}
S.~D. Ellis, C.~K. Vermilion, J.~R. Walsh, A.~Hornig, and C.~Lee,
\newblock JHEP {\bf 11}, 101 (2010), arXiv:1001.0014.

\bibitem{Blazey:2000qt}
G.~C. Blazey {\em et~al.},
\newblock {Run II jet physics},
\newblock in {\em {QCD and weak boson physics in Run II. Proceedings, Batavia,
  USA, March 4-6, June 3-4, November 4-6, 1999}}, pp. 47--77, 2000,
  arXiv:hep-ex/0005012.

\bibitem{Ellis:1993tq}
S.~D. Ellis and D.~E. Soper,
\newblock Phys. Rev. {\bf D48}, 3160 (1993), arXiv:hep-ph/9305266.

\bibitem{Salam:2007xv}
G.~P. Salam and G.~Soyez,
\newblock JHEP {\bf 05}, 086 (2007), arXiv:0704.0292.

\bibitem{Dokshitzer:1997in}
Y.~L. Dokshitzer, G.~D. Leder, S.~Moretti, and B.~R. Webber,
\newblock JHEP {\bf 08}, 001 (1997), arXiv:hep-ph/9707323.

\bibitem{Cacciari:2008gp}
M.~Cacciari, G.~P. Salam, and G.~Soyez,
\newblock JHEP {\bf 04}, 063 (2008), arXiv:0802.1189.

\bibitem{Kang:2017mda}
Z.-B. Kang, F.~Ringer, and W.~J. Waalewijn,
\newblock JHEP {\bf 07}, 064 (2017), arXiv:1705.05375.

\bibitem{Bai:2015fka}
Y.~Bai, Z.~Han, and R.~Lu,
\newblock (2015), arXiv:1509.07522.

\bibitem{Procura:2009vm}
M.~Procura and I.~W. Stewart,
\newblock Phys. Rev. {\bf D81}, 074009 (2010), arXiv:0911.4980,
\newblock [Erratum: Phys. Rev.D83,039902(2011)].

\bibitem{Procura:2011aq}
M.~Procura and W.~J. Waalewijn,
\newblock Phys. Rev. {\bf D85}, 114041 (2012), arXiv:1111.6605.

\end{thebibliography}
\end{document}